\documentclass[11pt,a4paper]{article}
\pdfoutput=1

\usepackage{jheppub}
\usepackage{color}
\usepackage{graphicx}
\usepackage{wrapfig,enumerate,slashed}

\usepackage{footmisc}
\usepackage{amsmath}
\usepackage{wasysym} 
\usepackage{graphicx}
\usepackage{color}
\usepackage{comment}
\usepackage{hyperref}
\usepackage{epstopdf}
\usepackage{caption}
\usepackage{subcaption}

\newcommand{\be}{\begin{eqnarray}}
\newcommand{\ee}{\end{eqnarray}}

\newcommand{\bee}{\begin{eqnarray}}
\newcommand{\eee}{\end{eqnarray}}
\newcommand{\beeq}{\begin{equation}}
\newcommand{\eeeq}{\end{equation}}

\renewcommand{\vec}{\bf}

\newcommand{\kp}{\kappa_{+}}
\newcommand{\km}{\kappa_{-}}

\newcommand{\GeV}{\text{~GeV}}

\newcommand{\Tr}[1]{\text{Tr}\left[#1\right]}

\graphicspath{{Figures/}}

\begin{document}

\title{Prospects for new physics in \boldmath $\tau \to l \mu \mu$ at current and future colliders}

\author[a]{Chris Hays,}
\author[b,c]{Manimala Mitra,}
\author[c]{Michael Spannowsky,}
\author[c]{and Philip Waite}

\affiliation[a]{Department of Physics, Oxford University, Oxford, OX1 3RH, UK}
\affiliation[b]{Department of Physics, Indian Institute of Science Education and Research Mohali (IISER Mo- hali), Sector 81, SAS Nagar, Manauli 140306, India}
\affiliation[c]{Institute for Particle Physics Phenomenology, Department of Physics, Durham University, Durham, DH1 3LE, UK}

\emailAdd{chris.hays@physics.ox.ac.uk}
\emailAdd{manimala@iisermohali.ac.in}
\emailAdd{michael.spannowsky@durham.ac.uk}
\emailAdd{p.a.waite@durham.ac.uk}

\abstract{The discovery of lepton flavour violating interactions will be striking evidence for physics 
beyond the Standard Model. Focusing on the three decays $\tau^{\mp} \to \mu^{\pm}\mu^{\mp}\mu^{\mp}$, 
$\tau^\mp \to e^\pm \mu^\mp \mu^\mp $ and $\tau^\mp \to e^\mp \mu^\mp \mu^\pm $, we evaluate the 
discovery potential of current and future high-energy colliders to probe lepton flavour violation in 
the $\tau$ sector.  Based on this potential we determine the expected constraints on parameters of new 
physics in the context of the Type-II Seesaw Model, the Left-Right Symmetric Model, and the Minimal 
Supersymmetric Standard Model. The existing and ongoing 13 TeV run of the Large Hadron Collider has the potential to produce constraints that outperform the existing $e^+ e^-$ collider 
limits for the $\tau^{\mp} \to \mu^{\pm}\mu^{\mp}\mu^{\mp}$ decay and achieve a branching fraction limit of $\lesssim 10^{-8}$.  With 
a future circular $e^+ e^-$ collider, constraints on the $\tau \to l \mu\mu$ branching fractions could reach 
as low as a few times $10^{-12}$.}

\keywords{Beyond Standard Model, Gauge Symmetry, Higgs Physics, Neutrino Physics}

\preprint{IPPP/16/126}

\maketitle


\section{Introduction}
\label{sec:intro}
In the Standard Model, the Yukawa couplings break the global flavour group $G_F$ explicitly to an accidental subgroup $G_F \equiv SU(3)^5 \to U(1)_B \times U(1)_{L_1} \times U(1)_{L_2}\times U(1)_{L_3}$. Hence, the model exhibits flavour conservation to all orders in perturbation theory that prohibits any process where charged lepton flavour is not conserved. Despite the immense success of the Standard Model (SM), it does not  serve  as an adequate description of nature due to its inability to explain  the  experimentally observed non-zero neutrino masses and mixings, the  radiative stability of the Higgs mass, and the existence of dark matter, for which beyond the Standard Model (BSM) descriptions are necessary.  Going beyond the SM, the models  that successfully explain the above problems often introduce lepton flavour violation (LFV) either at tree-level or via loop-induced processes. 

A selection of the interesting models that provide large lepton flavour violation are the various seesaw models \cite{Minkowski:1977sc, Mohapatra:1979ia, Yanagida:1979as,GellMann:1980vs,Schechter:1980gr,Magg:1980ut, Lazarides:1980nt, Cheng:1980qt, Mohapatra:1980yp,Foot:1988aq,Mohapatra:1986aw, Mohapatra:1986bd, Wyler:1982dd, Witten:1985bz, Hewett:1988xc},  the Left-Right Symmetric Model (LRSM) \cite{Pati:1974yy,Mohapatra:1974gc,Senjanovic:1975rk,Duka:1999uc}, and the Minimal Supersymmetric Standard Model (MSSM) \cite{Nilles:1983ge, Haber:1984rc, Martin:1997ns}. In the seesaw framework,  small Majorana masses of the light neutrinos are generated from the dimension-5 operator ${LLHH}/{\Lambda}$ \cite{Weinberg:1979sa, Wilczek:1979hc} through electroweak symmetry breaking. The high-scale theory  contains a plethora of  new particles, such as an extended neutrino  sector for the Type-I \cite{Minkowski:1977sc, Mohapatra:1979ia, Yanagida:1979as, GellMann:1980vs, Schechter:1980gr}, Type-III \cite{Foot:1988aq}  and inverse seesaw \cite{Mohapatra:1986aw, Mohapatra:1986bd, Wyler:1982dd, Witten:1985bz, Hewett:1988xc} models, and an extended scalar sector for the Type-II Seesaw Model \cite{Magg:1980ut,Lazarides:1980nt}. In the LRSM \cite{Pati:1974yy,Mohapatra:1974gc,Senjanovic:1975rk,Duka:1999uc}, the model contains both extended neutrino and Higgs sectors, and the light neutrino masses are generated via a combination of Type-I and Type-II seesaw mechanisms. The non-trivial interactions of the heavy neutrinos or scalars with the SM  charged leptons allow for a priori unsuppressed LFV interactions in these theories. In the MSSM, the large LFV is introduced by the non-diagonal slepton mass matrices.  The large LFV rates of these new particles can be tested at present and future colliders. Hence, experimental evidence for a non-zero LFV rate  will serve as striking evidence for the existence of physics beyond the Standard Model. 

The existing experimental constraints for LFV in transitions between the first and second generations are 
quite tight: $\mathrm{BR}(\mu \to e \gamma)  \leq 5.7 \times 10^{-13}$ at 90$\%$ confidence level (C.L.) 
as reported by MEG \cite{Adam:2013mnn, Olive:2016xmw},  and $\mathrm{BR}(\mu^{\mp} \to e^{\pm} e^{\mp} e^{\mp}) \leq 10^{-12} $ at 
90$\%$ C.L. \cite{Bellgardt:1987du, Olive:2016xmw}.  Lepton flavour violation in $\tau$ lepton decays is 
much less constrained: $\mathrm{BR}(\tau \to lll )  \lesssim  10^{-8}$ at 90$\%$ C.L. \cite{Olive:2016xmw}, 
allowing for rather large flavour violating couplings. Considering  low-energy models, one can avoid the 
stringent constraints from the LFV processes involving the first and second generations. The recent excess in $h\to \tau \mu$ reported by 
CMS~\cite{Khachatryan:2015kon}, as well as a smaller excess by ATLAS~\cite{Aad:2015gha}, spurred further 
interest in collider studies of flavour-changing neutral interactions in decays of the Higgs boson and the 
$\tau$ lepton \cite{Dassinger:2007ru, Harnik:2012pb,Falkowski:2013jya,Heeck:2014qea,Crivellin:2015mga,Banerjee:2016foh,Chakraborty:2016gff}. 
Experimental limits from the Belle and BaBar experiments at the flavour factories are currently the most 
stringent, requiring the branching ratio for the $\tau^{\mp} \to \mu^{\pm} \mu^{\mp} \mu^{\mp}$ decay to 
be less than $2.1 \times 10^{-8}$ at $90\%$ C.L.  Similar exclusion limits are obtained for the 
$\tau^{\mp} \to e^{\pm} \mu^{\mp} \mu^{\mp} $ and  $\tau^{\mp} \to e^{\mp} \mu^{\mp} \mu^{\pm} $ modes.  
A recent search from the LHCb experiment produced a competitive constraint for the 
$\tau^{\mp} \to \mu^{\pm} \mu^{\mp} \mu^{\mp}$ decay, with the limit a factor of two larger than the constraints 
from Belle, $\textrm{BR}(\tau^{\mp} \to \mu^{\pm} \mu^{\mp} \mu^{\mp}) \leq 4.6\times 10^{-8}$ at 90$\%$ C.L.  
The recent bound from ATLAS is one order of magnitude smaller, though current and future 13 TeV data sets from 
ATLAS and CMS can significantly extend this sensitivity to 
$\textrm{BR}(\tau^{\mp} \to \mu^{\pm} \mu^{\mp} \mu^{\mp}) \sim 10^{-9}$. 
The Belle-II experiment and a possible future circular collider will be sensitive to even lower branching ratios,   
$\sim 10^{-10}$ and $\sim 10^{-12}$, respectively. 

In this work we analyse LFV in the $\tau$ sector, focusing on  the decay modes $\tau^{\mp} \to \mu^{\pm}\mu^{\mp}\mu^{\mp}$,  $\tau^\mp \to e^\pm \mu^\mp \mu^\mp $, and  $\tau^\mp \to e^\mp \mu^\mp \mu^\pm$. We  consider the potential of both 
$e^+e^-$ and hadron colliders, including future circular colliders, in searching for LFV in $\tau$ lepton decays.  Using the expected constraints we derive the sensitivity reach for three BSM models: the Type-II Seesaw Model, the LRSM, and the MSSM. 

The rest of the paper is organised as follows:  In Sec.~\ref{sec:limits}, we discuss current and future limits from flavour factories and high-energy colliders on rare flavour violating $\tau$ decays.  In Sec.~\ref{sec:models},  we test popular and widely studied extensions of the SM, such as the Type-II Seesaw Model, the LRSM and the MSSM, using the limits collected in Sec.~\ref{sec:limits}. While the Type-II Seesaw Model and the LRSM induce tree-level LFV interactions,  LFV processes are generically loop suppressed in the MSSM. Nonetheless, particularly for the former two models \cite{Hirsch:1996qw,Barry:2013xxa,Awasthi:2013ff,Bambhaniya:2015ipg,Bonilla:2016fqd} but also for the MSSM \cite{Arganda:2008jj,Arana-Catania:2013ggc, Arana-Catania:2013xma}, LFV has become a litmus test, excluding large areas of the parameter space. Finally, in Sec.~\ref{sec:conclusions}, we present our conclusions. 

\section{Experimental limits}
\label{sec:limits}
We review present and future collider constraints on the processes $\tau^{\mp} \to \mu^{\pm} \mu^{\mp} \mu^{\mp}$, 
$\tau^\mp \to e^\pm \mu^\mp \mu^\mp$ and $\tau^\mp \to e^\mp \mu^\mp \mu^\pm$.  
Limits on $\tau$ lepton decays to three charged leptons have been obtained at both $e^+ e^-$ and 
hadron colliders, with the B-factories currently giving the most stringent limits.  However, the data from 
the LHC run at $\sqrt{s}=13$~TeV could result in stronger $\tau^{\mp} \to \mu^{\pm} \mu^{\mp} \mu^{\mp}$ limits than those from 
B-factories.  In the long run, the upgraded KEKB $e^+ e^-$ collider and a potential future circular $e^+ e^-$ 
collider are expected to provide the greatest sensitivity to these processes.

\subsection{Current limits}

The Belle and BaBar experiments probe the six possible combinations of $\tau$ lepton decays to three 
charged leptons using $e^+ e^-$ integrated luminosities of 782 fb$^{-1}$~\cite{Hayasaka:2010np} and 
468 fb$^{-1}$~\cite{babar}, respectively, representing nearly the complete available data sets.  
The $\tau^+ \tau^-$ cross section is 0.919~nb, giving 720 (430) million $\tau$ lepton pairs in 
the Belle (BaBar) data set.  Events are selected at Belle by requiring one identified $\tau$ lepton 
decay (the ``tag'' $\tau$ lepton) and searching for a lepton flavour violating $\tau$ lepton decay (the 
``signal'' $\tau$ lepton).  The background is very low after a basic selection and is primarily 
due to $\tau^+ \tau^-$ production or quark-antiquark production with misidentified leptons for the 
$\tau^{\mp} \to \mu^{\pm} \mu^{\mp} \mu^{\mp}$ and $\tau^\mp \to e^\pm \mu^\mp \mu^\mp$ searches.  
For the $\tau^\mp \to e^\mp \mu^\mp \mu^\pm$ decay the main contribution is $\gamma\gamma \to \mu^+ \mu^-$ 
with a scattered electron.  In the $\tau^{\mp} \to \mu^{\pm} \mu^{\mp} \mu^{\mp}$ case, an additional 
background rejection is applied using the missing momentum and missing mass-squared in the event.  
This decreases the efficiency of the selection to 7.6\% (the efficiency of the 
$\tau^\mp \to e^\pm \mu^\mp \mu^\mp$ selection is 10.1\%).  The expected background, estimated from 
data, is $0.02-0.13$ events.  No events are observed and the 90\% C.L. upper 
limits on the branching fractions are 
$2.1\times 10^{-8}$, $1.7\times 10^{-8}$, and $2.7\times 10^{-8}$ for $\tau^{\mp} \to \mu^{\pm} \mu^{\mp} \mu^{\mp}$, 
$\tau^\mp \to e^\pm \mu^\mp \mu^\mp$, and $\tau^\mp \to e^\mp \mu^\mp \mu^\pm$ respectively.
The corresponding limits from BaBar are $3.3 \times 10^{-8}$, $2.6 \times 10^{-8}$, and $3.2 \times 10^{-8}$.

The LHCb experiment has searched for $\tau^{\mp} \to \mu^{\pm} \mu^{\mp} \mu^{\mp}$ in 3~fb$^{-1}$ of $pp$ collision 
data at centre-of-mass energies of 7 and 8 TeV~\cite{Aaij:2014azz}.  The production of $\tau$ 
leptons at the LHC occurs predominantly through the decays of heavy quarks, with an inclusive cross 
section of approximately $85~\mu$b.  The $\tau$ lepton yield is normalised using the 
$D_s \to \phi(\mu\mu)\pi$ decay, the relative branching fractions for 
$D_s \to \phi(\mu\mu)\pi$ and $D_s \to \tau\nu$, and the fraction of $\tau$ leptons 
that are produced via $D_s \to \tau\nu$.  Backgrounds from 
$D_s \to \eta(\mu\mu\gamma)\mu\nu$ decays motivate a fit of the three-muon mass 
distribution in 30 (35) bins of particle-identification and geometric-event classifiers in 
$\sqrt{s}=7$~(8)~TeV data.  The fit describes the background as an exponential distribution in the 
mass range $(1600-1950)$~MeV, excluding the signal window of $\pm 30$~MeV around the $\tau$ lepton 
mass.  The observed yields in the signal region are consistent with the background and range from 
0 to 39 events, with the highest yields present in bins of the particle identification classifier 
where the misidentification backgrounds $D_{(s)} \to K\pi\pi$ and 
$D_{(s)} \to \pi\pi\pi$ are significant.  These bins are excluded when deriving the 
90\% C.L. upper limit of $4.6 \times 10^{-8}$ on the branching fraction for 
$\tau^{\mp} \to \mu^{\pm} \mu^{\mp} \mu^{\mp}$.

Finally, the ATLAS experiment has searched for $\tau^{\mp} \to \mu^{\pm} \mu^{\mp} \mu^{\mp}$ decays using 8~TeV 
$pp$ collision data corresponding to an integrated luminosity of 20.3~fb$^{-1}$~\cite{Aad:2016wce}.  
The search selects candidate $W$ boson decays using the missing transverse momentum 
($\vec{p}_{\rm T}^{\rm miss}$) and the transverse mass 
$m_{\rm T} = \sqrt{2p_{\rm T}^{\tau} p_{\rm T}^{\rm miss}(1-\cos\Delta\phi)}$, 
where $\Delta\phi$ is the angle between $\vec{p}_{\rm T}^{\tau}$ and $\vec{p}_{\rm T}^{\rm miss}$.
Candidate lepton flavour violating decays are defined as those with three muons within 1 GeV of the 
mass of the $\tau$ lepton, and a loose selection is applied based on kinematics and displacement 
of the three-muon vertex relative to the collision point.  The large multi-jet background is then 
removed using a boosted decision tree (BDT) and requiring the three-muon mass to be within 
$\pm 64$~MeV of the $\tau$ lepton mass.  The optimal BDT selection leaves 0.2 expected background 
events with an efficiency of 2.3\%.  No events are observed, leading to a 90\% C.L. upper limit of 
$3.8 \times 10^{-7}$ on the branching fraction.  

\subsection{Future limits}

\begin{table}[!tb]
\begin{center}
\begin{tabular}{lll}
\hline
\hline
Experiment & Current & Projected \\
  \hline
Belle  & $2.1\times 10^{-8}$ & $(4.7-10) \times 10^{-10}$ \\
BaBar  & $3.3 \times 10^{-8}$ & \multicolumn{1}{c}{$-$} \\
FCC-ee & \multicolumn{1}{c}{$-$} & $(5-10) \times 10^{-12}$ \\
LHCb   & $4.6 \times 10^{-8}$ & $(1.5-11) \times 10^{-9}$ \\ 
ATLAS  & $3.8 \times 10^{-7}$ & $(1.8-8.1) \times 10^{-9}$ \\
FCC-hh & \multicolumn{1}{c}{$-$} & $(3-30) \times 10^{-10}$ \\
\hline
\hline
\end{tabular}
\end{center}
\caption{Current and projected 90\% C.L. limits on the $\tau^{\mp} \to \mu^{\pm} \mu^{\mp} \mu^{\mp}$ 
branching fraction.  The current limits from the LHC experiments utilise only the 8 TeV data, while 
the projected limits are based on the complete 13 TeV data sets of 3~ab$^{-1}$ for ATLAS and 50~fb$^{-1}$ 
for LHCb from the high-luminosity run of the LHC. }
\label{tab:expsummary3mu}
\end{table}

Projections of the current analyses are complicated by the prevalence of misidentification backgrounds, 
which typically require data to model.  A conservative estimate scales the background yield by the 
projected increase in luminosity and cross section.  However, further optimisation of the analyses 
incorporating upgrades to the detectors could improve these results.  As an optimistic estimate the 
background is kept at the current level with a modest 10\% loss of acceptance.  

An ongoing upgrade to the KEK accelerator and the Belle detector (Belle-II) will ultimately yield a factor of 
50 increase in integrated luminosity, with data taking set to begin in 2017.  A conservative 
estimate of the expected $\tau^{\mp} \to \mu^{\pm} \mu^{\mp} \mu^{\mp}$ sensitivity can be made by simply scaling the 
background from 0.13 to 6.5 events and assuming no change in the reconstruction efficiency.  This 
leads to an expected upper limit of $1.0 \times 10^{-9}$ on the branching fraction (equal to the 
projected limit from the experiment~\cite{belleprojection}).  Including a more optimistic projection, 
the ranges of expected limits are $(4.7-10) \times 10^{-10}$, $(3.6-4.7) \times 10^{-10}$, and 
$(5.9-12)\times 10^{-10}$ on the branching fractions for $\tau^{\mp} \to \mu^{\pm} \mu^{\mp} \mu^{\mp}$, 
$\tau^\mp \to e^\pm \mu^\mp \mu^\mp$, and $\tau^\mp \to e^\mp \mu^\mp \mu^\pm$, respectively.

The upgrade of the LHC accelerator and the LHCb detector will produce a data sample corresponding to 
an integrated luminosity of 50 fb$^{-1}$~\cite{Bediaga:2012py} at a centre-of-mass energy of 13 TeV.  
Taking the ratio of 13 TeV to 7 TeV heavy-quark production cross section to be 
1.8~\cite{Aaij:2011jh, Aaij:2015rla, Aaij:2013mga,Aaij:2015bpa}, the $\tau$ lepton yield will increase 
by approximately a factor of 30.  Taking into account the higher background cross section, a 
conservative estimate of the expected limit is $1.1 \times 10^{-8}$.  A more optimistic estimate 
assuming the background can be reduced to its current level gives a 90\% C.L. upper limit of 
$1.5 \times 10^{-9}$ on the $\tau^{\mp} \to \mu^{\pm} \mu^{\mp} \mu^{\mp}$ branching fraction.

\begin{table}[!tb]
\begin{center}
\begin{tabular}{lllll}
\hline
\hline
 & \multicolumn{2}{c}{$\tau^\mp \to e^\pm \mu^\mp \mu^\mp$} & \multicolumn{2}{c}{$\tau^\mp \to e^\mp \mu^\mp \mu^\pm$} \\
Experiment & Current & Projected & Current & Projected \\
  \hline
Belle  & $1.7\times 10^{-8}$ & $(3.4-5.1)\times 10^{-10}$ & $2.7 \times 10^{-8}$ & $(5.9-12)\times 10^{-10}$ \\
BaBar  & $2.6\times 10^{-8}$ & \multicolumn{1}{c}{$-$} & $3.2 \times 10^{-8}$ & \multicolumn{1}{c}{$-$} \\
FCC-ee & \multicolumn{1}{c}{$-$} & $(5-10) \times 10^{-12}$ & \multicolumn{1}{c}{$-$} & $(5-10) \times 10^{-12}$ \\
\hline
\hline
\end{tabular}
\end{center}
\caption{Current and projected 90\% C.L. limits on the $\tau^\mp \to e^\pm \mu^\mp \mu^\mp$ and 
$\tau^\mp \to e^\mp \mu^\mp \mu^\pm$ branching fractions.}
\label{tab:expsummarye2mu}
\end{table}

The ATLAS sensitivity to the high-luminosity LHC will be affected by a high number of overlapping 
interactions, potentially leading to lower neutrino momentum resolution and lower trigger efficiencies.  
Assuming the current performance is approximately achieved through detector upgrade and analysis 
improvements, the expected $\tau$ lepton yields can be scaled to 3~ab$^{-1}$ with a factor of 1.6 
increase in cross section~\cite{Chatrchyan:2014mua,Aad:2016naf}.  Assuming an equal scaling for the 
background gives 46 expected background events and a 90\% C.L. of $8.1 \times 10^{-9}$.  In the most 
optimistic scenario, where the background is suppressed to its current level with a modest 10\% 
efficiency loss, the expected 90\% C.L. on the $\tau^{\mp} \to \mu^{\pm} \mu^{\mp} \mu^{\mp}$ branching 
fraction is $1.8 \times 10^{-9}$.

A future circular collider (FCC)~\cite{FCC} could further improve sensitivity to these processes.  A 
proton-proton collider with $\sqrt{s}=100$~TeV would have $\sim 7$ times the cross section for $W$ 
and $Z$ boson production than the LHC~\cite{Mangano:2016jyj}.  Assuming a detector with equivalent 
sensitivity to ATLAS, projecting the conservative and optimistic limits to 3~ab$^{-1}$ of integrated 
luminosity of a 100 TeV collider gives a range of $(3-30) \times 10^{-10}$ for the 90\% C.L. on the 
$\tau^{\mp} \to \mu^{\pm} \mu^{\mp} \mu^{\mp}$ branching fraction.  Better sensitivity could be achieved 
by an $e^+ e^-$ collider producing 55~ab$^{-1}$ of integrated luminosity on the $Z$ resonance at four 
interaction points~\cite{dEnterria:2016sca}.  Such a collider would produce a total of $\sim 6 \times 10^{11}$ 
$\tau$ leptons, and a typical detector could identify rare decays with a high efficiency and low 
background.  Taking an efficiency of $(40-80)\%$ and the background to be negligible, 90\% C.L. upper 
limits would range from $(5-10) \times 10^{-12}$ on the branching fractions for all lepton flavour 
violating $\tau$ lepton decays.  Given the high potential sensitivity of such a collider, a more 
careful assessment is warranted.

In summary, the strongest present limits on $\tau^{\mp} \to \mu^{\pm} \mu^{\mp} \mu^{\mp}$ come from 
Belle and will improve by an order of magnitude to $\leq 10^{-9}$ with the expected 50-fold increase in 
luminosity from SuperKEKB.  Constraints from the LHCb and ATLAS experiments could be within a factor 
of two of these limits.  If CMS can provide similar sensitivity, then the combined hadron collider results 
could exceed the sensitivity of the $e^+ e^-$ constraints.  Further gains are possible at the LHC if 
decays of heavy-flavour mesons and $W$ and $Z$ bosons can all be used by the experiments.  In the short 
term, with the 2016 and 2017 data the LHC experiments could overtake the current Belle and BaBar limits.  In 
the far future, a circular $e^+ e^-$ collider with a centre-of-mass energy on the $Z$ resonance could further 
improve constraints by two orders of magnitude.  Table~\ref{tab:expsummary3mu} summarises the current 
and projected limits on the $\tau^{\mp} \to \mu^{\pm} \mu^{\mp} \mu^{\mp}$ branching fraction, and 
Table~\ref{tab:expsummarye2mu} shows the equivalent limits for $\tau^\mp \to e^\pm \mu^\mp \mu^\mp$ and 
$\tau^\mp \to e^\mp \mu^\mp \mu^\pm$.

\section{Standard Model extensions with lepton flavour violating interactions}
\label{sec:models}

\begin{figure}[!tb]
\begin{subfigure}[b]{0.3\textwidth}
\includegraphics[width=\textwidth]{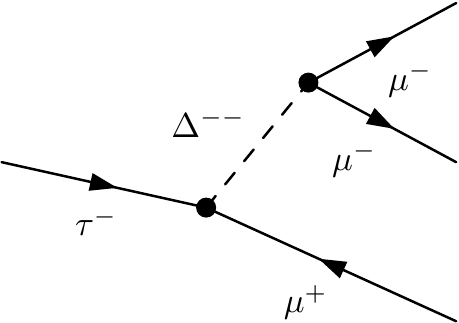}
\caption{}
\label{fig:T2csaw-tau_3mu-diagram}
\end{subfigure}
\hspace{0.5cm}
\begin{subfigure}[b]{0.3\textwidth}
\includegraphics[width=\textwidth]{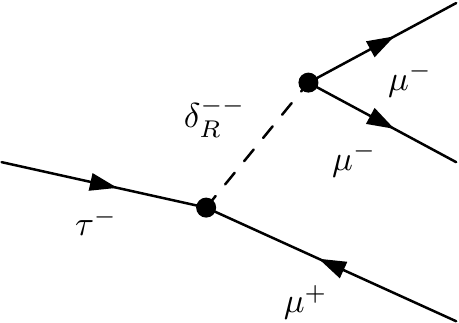}
\caption{}
\label{fig:MLRSM-tau_3mu-diagram}
\end{subfigure}
\hspace{0.5cm}
\begin{subfigure}[b]{0.3\textwidth}
\includegraphics[width=\textwidth]{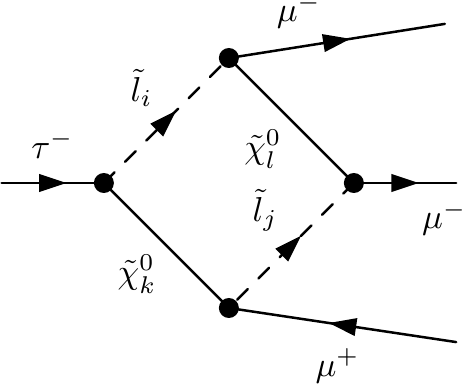}
\caption{}
\label{fig:mssm-tau_3mu-diagram}
\end{subfigure}
\caption{Characteristic Feynman diagrams for the decay $\tau^{\mp} \to \mu^{\pm} \mu^{\mp} \mu^{\mp}$ in (a) the Type-II Seesaw Model, (b) the LRSM and (c) the MSSM.}
\label{fig:feynman}
\end{figure}

Following the effective field theory (EFT) approach, lepton flavour violating interactions $l_i \to l_j l_k l_l$ can be induced
via the dimension-6 operators ${\mathcal{\hat{O}}_6}=c_{ijkl}\,l_i l_j l_k l_l/ \Lambda^2$. These LFV operators are generated  from the high-scale BSM theories once  the heavy particles of the BSM theory are integrated out. As the prototype examples, in the following subsections we consider three BSM extensions: the Type-II Seesaw Model, the Left-Right Symmetric Model and the Minimal Supersymmetric Standard Model. It is worth noting that the chosen seesaw models can generate large LFV rates $l_i \to l_j l_k l_l$ at tree-level and hence can be highly constrained by the present and future LFV searches. For the MSSM, large flavour violation arises at a loop-induced level. An example Feynman diagram for the process $\tau^{\mp} \to \mu^{\pm} \mu^{\mp} \mu^{\mp}$  for each model is shown in Fig.~\ref{fig:feynman}. For the computations of the branching ratios in the Type-II Seesaw Model and the LRSM, we use the program \texttt{MadGraph5\_{}aMC@NLO} \cite{Alwall2014} with the model files generated by \texttt{FeynRules} \cite{Alloul20142250}. For the loop-induced decays in the MSSM, we use the spectrum generator \texttt{SPheno} \cite{POROD2003275, Porod20122458}, with the source code for the flavour observables produced by \texttt{SARAH} \cite{Staub20141773}. We note that the BSM 
particles that produce this indirect signature could also be directly produced at colliders.  For a recent discussion on 
the collider studies of the seesaw models, see~\cite{ATLAS-CONF-2016-051, Aad:2015xaa, Khachatryan:2014dka, ATLAS:2015nsi, Khachatryan:2015dcf,  Deppisch:2015qwa, Babu:2016rcr, Nemevsek:2011hz, Nemevsek:2016enw, Mitra:2016wpr, Mitra:2016kov, Mattelaer:2016ynf, Lindner:2016lxq, Lindner:2016lpp,  Melfo:2011nx,Maiezza:2015lza, delAguila:2008cj, Atre:2009rg, Dev:2016dja, Banerjee:2015gca, Dev:2013wba,Das:2012ze,Das:2016hof}. 



\subsection{Type-II Seesaw Model}
The model consists of the SM Higgs doublet $\Phi$ supplemented by an additional Higgs triplet $\Delta$ with hypercharge $Y=+2$,  
\begin{eqnarray}
\Phi= \begin{pmatrix} \Phi^+ \\ \Phi^0  \end{pmatrix}~,
\qquad
\Delta=\begin{pmatrix} \frac{\Delta^+}{\sqrt{2}} & \Delta^{++} \\ \Delta^0 & -\frac{\Delta^+}{\sqrt{2}}
\end{pmatrix}~.
\end{eqnarray}
The neutral component $\Delta^0$ has the vacuum expectation value (vev) $v_{\Delta}$, and generates the Majorana masses of the light neutrinos $M_{\nu}$.  
The interaction of $\Delta$ with the two lepton doublets is given by, 
\begin{eqnarray}
\mathcal{L}_Y(\Phi, \Delta)&=& Y_{\Delta}\overline{L_L^{c}}i\tau_2\Delta L_L+\mathrm{h.c.}~.
\label{yukawa}
\end{eqnarray}
Here, $c$ denotes the charge conjugation transformation $\tilde{\Phi}=i \sigma_2 \Phi^{*}$, while $Y_{\Delta}$ is the Yukawa matrix. The light neutrino mass matrix is proportional to the vev $v_{\Delta}$, with 
\be
M_{\nu}=\sqrt{2}Y_{\Delta} v_{\Delta}~, 
\label{t2seesaw}
\ee 
where the triplet vev $v_{\Delta}$ is $ v_{\Delta}=\mu_{\Delta} v^2_{\Phi}/(\sqrt{2} M^2_{\Delta})$, and $v_{\Phi}$ is the electroweak vev. We note that an equivalent description of the Type-II seesaw is with the triplet Higgs field $\Delta$ that  gets integrated out and generates the dimension-5 operator $L_iL_jHH/\Lambda$ with  the coefficient $C_{ij} = Y_{\Delta} \mu_{\Delta}/M^2_{\Delta}$. 
The Yukawa Lagrangian generates the following interaction terms between the doubly charged Higgs field $\Delta^{++}$ and the pairs of leptons
($\mu$, $\tau$) and ($\mu,\mu$):
\begin{eqnarray}
\mathcal{L}_Y(\Delta^{++})&=& Y_{\mu \tau}\overline{\mu^{c}}\tau \Delta^{++}+ Y_{\mu \mu}\overline{\mu^{c}}\mu \Delta^{++}+\mathrm{h.c.}~.
\label{yukawadc}
\end{eqnarray}
In addition to the Yukawa Lagrangian, the Higgs triplet $\Delta$ interacts with the SM Higgs and gauge bosons through the scalar
potential and the kinetic Lagrangian. For a complete description of the scalar potential and the other interactions, see \cite{Arhrib:2011uy}. The 
 trilinear interaction of the $\Delta$ with the SM Higgs doublet is governed by the following Lagrangian:
\begin{eqnarray}
V(\Phi,\Delta)&=& \mu_{\Delta} \Phi^{\mathrm{T}}i\tau_2\Delta^\dagger \Phi+\mathrm{h.c.}~.
\label{eqn:scalpttri}
\end{eqnarray}
The Higgs triplet $\Delta$ carries  lepton number $+2$. The  simultaneous presence of  $Y_{\Delta}$ and $\mu_{\Delta}$ gives rise to lepton number violation in this model, while the off-diagonal elements in $Y_{\Delta}$ give rise to flavour violation. 

\begin{figure}[!tb]
\centering
\begin{subfigure}[b]{0.49\textwidth}
\centering
\includegraphics[width=\textwidth]{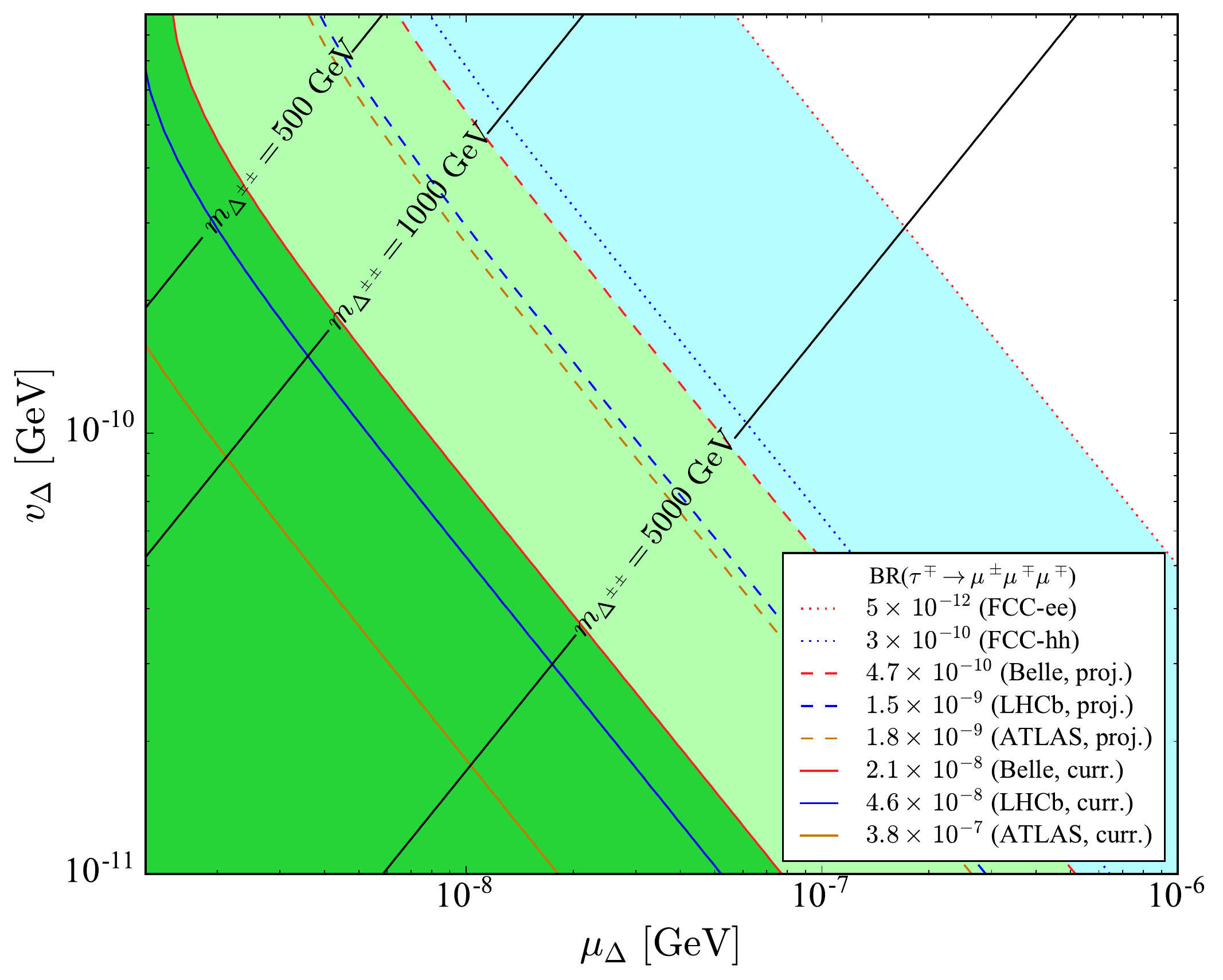}
\caption{}
\label{fig:T2csaw-tau_3mu}
\end{subfigure}
\hfill
\begin{subfigure}[b]{0.49\textwidth}
\centering
\includegraphics[width=\textwidth]{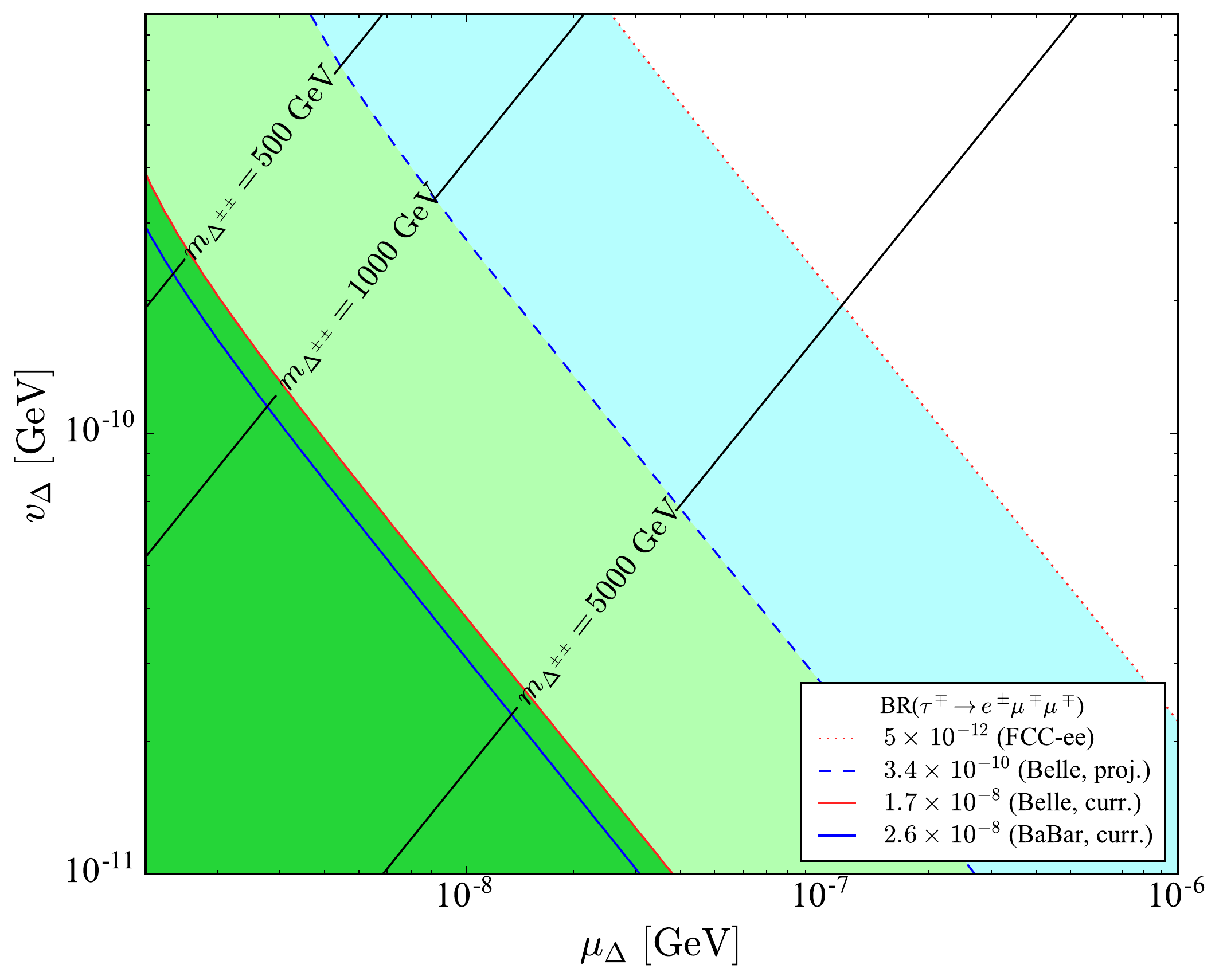}
\caption{}
\label{fig:T2csaw-tau_e2mu}
\end{subfigure}
\hfill
\centering
\begin{subfigure}[b]{0.49\textwidth}
\centering
\includegraphics[width=\textwidth]{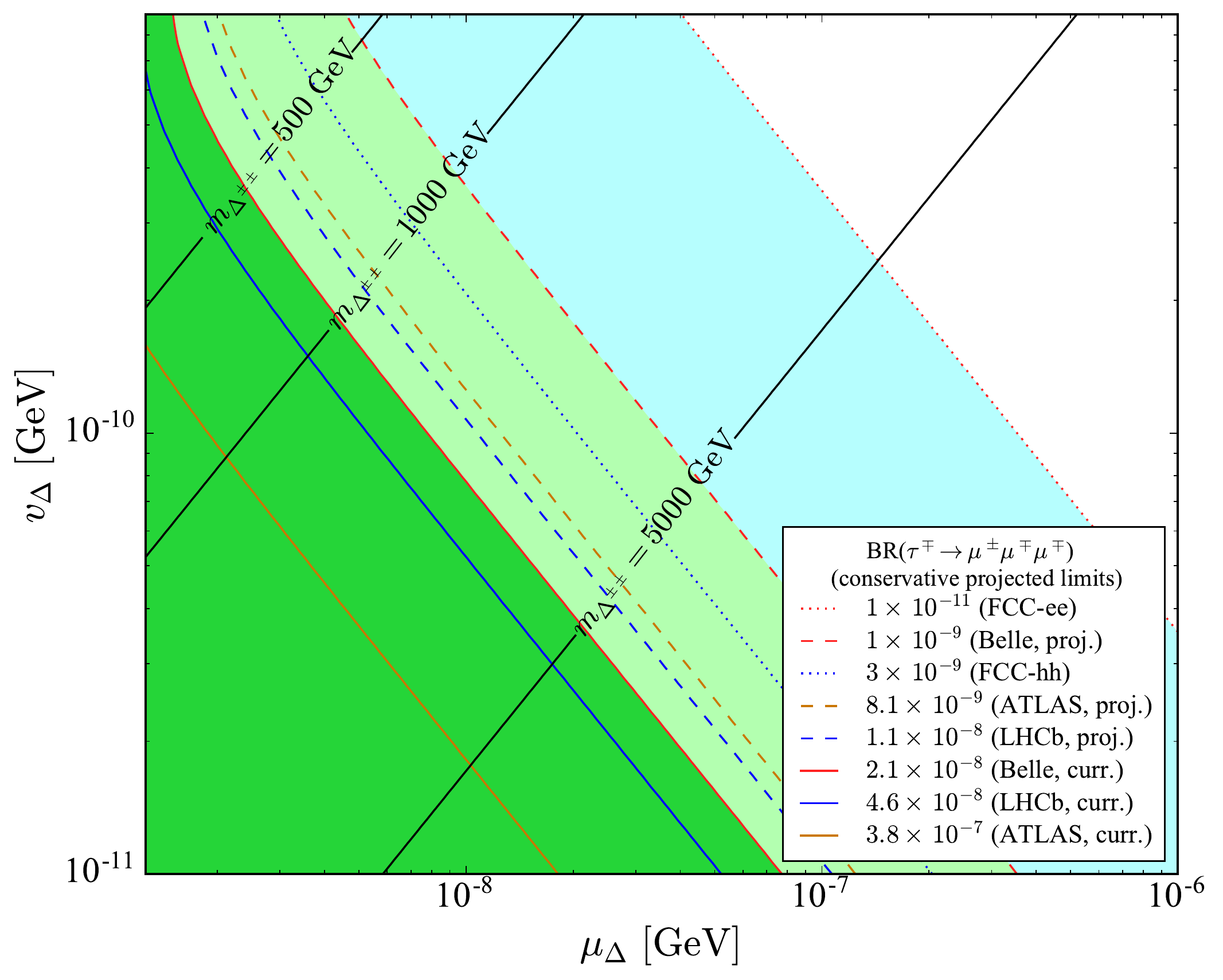}
\caption{}
\label{fig:T2csaw-tau_3mu-conservative}
\end{subfigure}
\hfill
\begin{subfigure}[b]{0.49\textwidth}
\centering
\includegraphics[width=\textwidth]{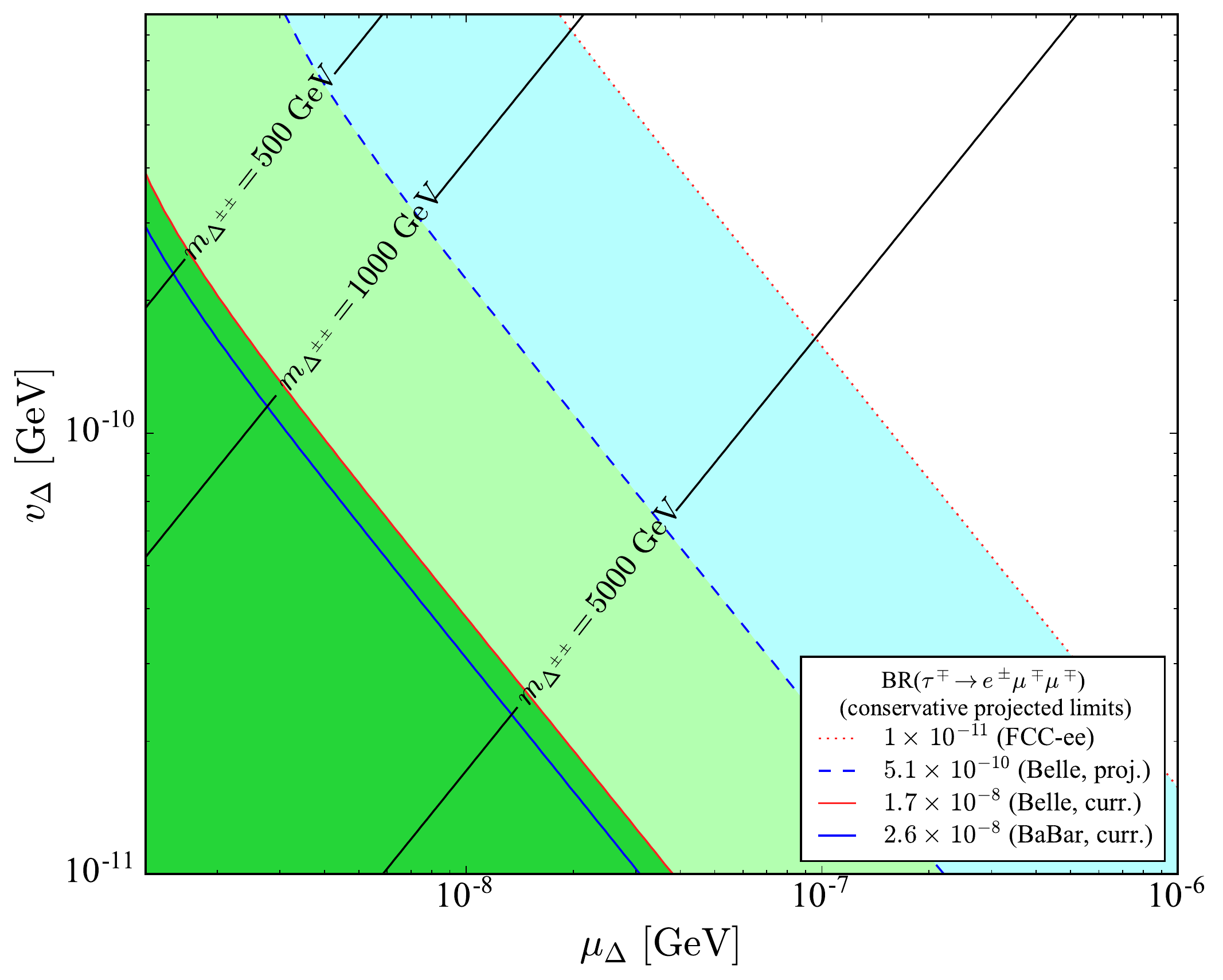}
\caption{}
\label{fig:T2csaw-tau_e2mu-conservative}
\end{subfigure}
\hfill
\caption{Current and future branching ratio limits in the parameter plane of $\mu_{\Delta}$ and $v_{\Delta}$ for the Type-II Seesaw Model. (a) Shows the limits from the decay $\tau^{\mp} \to \mu^{\pm} \mu^{\mp} \mu^{\mp}$, and (b) shows the limits from the decay $\tau^\mp \to e^\pm \mu^\mp \mu^\mp$. The same two decay processes are shown in (c) and (d) but with the conservative estimates for the projected limits instead. The solid black lines represent constant values of the mass of the doubly charged Higgs $\Delta^{\pm\pm}$.}
\label{fig:T2csaw_one}
\end{figure}

The interaction of the doubly charged Higgs with the two charged leptons gives rise to the lepton flavour violating Higgs decays $l_i \to l_j l_k l_l$. The  partial decay width for  $\tau^{\mp} \to \mu^{\pm} \mu^{\mp} \mu^{\mp}$ is given by \cite{Abada:2007ux},
\begin{eqnarray}
{\Gamma}(\tau^{\mp} \to \mu^{\pm} \mu^{\mp} \mu^{\mp}) &=& \frac{m^5_{\tau}}{192 \pi^3} |C_{\tau \mu \mu \mu}|^2~,  
\label{pwdtau3mu}
\end{eqnarray}
where the coefficient $C_{\tau \mu \mu \mu}$ has the following form:
\begin{eqnarray}
C_{ \tau \mu\mu\mu} &=& \frac{Y_{\tau \mu} Y_{\mu \mu}}{m^2_{\Delta^{\pm \pm}}}=\frac{M_{\nu}(\tau, \mu) M_{\nu}(\mu,\mu)}{2 v^2_{\Delta} m^2_{\Delta^{\pm \pm}}}~,
\label{pwdtau3mucoef}
\end{eqnarray}
where $m_{\Delta^{\pm \pm}}$ is the mass of the doubly charged Higgs and is given by, 
\begin{eqnarray}
\label{eq:mhpp}
m_{\Delta^{\pm \pm}}^2&=&M_\Delta^2-v_\Delta^2\lambda_3-\frac{\lambda_4}{2}v_\Phi^2~, \qquad M^2_{\Delta}=\frac{\mu_{\Delta} v^2_{\Phi}}{\sqrt{2}v_{\Delta}}~.
\end{eqnarray}
In the above, $\lambda_{3,4}$ are the couplings of the potential \cite{Arhrib:2011uy, Mitra:2016wpr}, and $v_{\Phi}$ is the vev of $\Phi$. 
The LFV rates for the process $\tau^\mp \to e^\pm \mu^\mp \mu^\mp$ can be obtained by replacing $M_{\nu}(\mu,\tau) $ with $M_{\nu}(e,\tau) $ in Eq.~\eqref{pwdtau3mu}. For detailed discussions on the LFV decays with the other bounds,  see \cite{Dinh:2013vya, Lindner:2016bgg, Akeroyd:2009nu,Chakrabortty:2015zpm}. Other LFV processes, such as $\mu^{\mp} \to e^{\pm} e^{\mp} e^{\mp}$, depend on a different combination of Yukawa couplings and can be suppressed for a large range of neutrino oscillation parameters and phases while still allowing for sizeable LFV $\tau$ lepton branching ratios. This was discussed in detail in \cite{Dinh:2013vya}, for both hierarchical and quasi-degenerate neutrino masses, where branching ratios of as large as $10^{-8}$ for $\tau^{\mp} \to \mu^{\pm} \mu^{\mp} \mu^{\mp}$ were obtained, while still being consistent with the other bounds. 
Here we focus on the bounds derived from the LFV $\tau$ lepton decays, independent of other constraints. At the end of this subsection, we will give a brief discussion of the consistency of our results with the other bounds when allowing for variations of the neutrino oscillation parameters and phases.

Fig.~\ref{fig:T2csaw_one} shows the current and future branching ratio limits in the plane of the parameters $\mu_{\Delta}$ and $v_{\Delta}$, for the two processes $\tau^{\mp} \to \mu^{\pm} \mu^{\mp} \mu^{\mp}$ and $\tau^\mp \to e^\pm \mu^\mp \mu^\mp$ respectively. We fix the neutrino masses and oscillation parameters to their best-fit values \cite{Esteban:2016qun, Gonzalez-Garcia:2015qrr} with the lightest neutrino mass at 0.1 eV, and take the PMNS phase to be zero. The solid black lines represent constant values of the doubly charged Higgs mass across the parameter plane. The dark green regions show the parameter space restricted by the  current limits, while the pale green regions show the exclusions that can be obtained by projections of current experiments. Furthermore, the pale blue regions show the restrictions from the future circular colliders FCC-hh and FCC-ee, while the white region is the part of the parameter space that will be allowed by the FCC-ee limit. For the projected limits we show the lower values of the limit ranges in Figs.~\ref{fig:T2csaw-tau_3mu} and \ref{fig:T2csaw-tau_e2mu}, corresponding to the best possible sensitivity for each experiment. In Figs.~\ref{fig:T2csaw-tau_3mu-conservative} and \ref{fig:T2csaw-tau_e2mu-conservative}, we instead show the most conservative estimates for the limits. All other parameter plots in this paper will follow the same scheme for the region colours, and will use the lower values of the limit ranges.

\begin{figure}[!tb]
\centering
\begin{subfigure}[b]{0.49\textwidth}
\centering
\includegraphics[width=\textwidth]{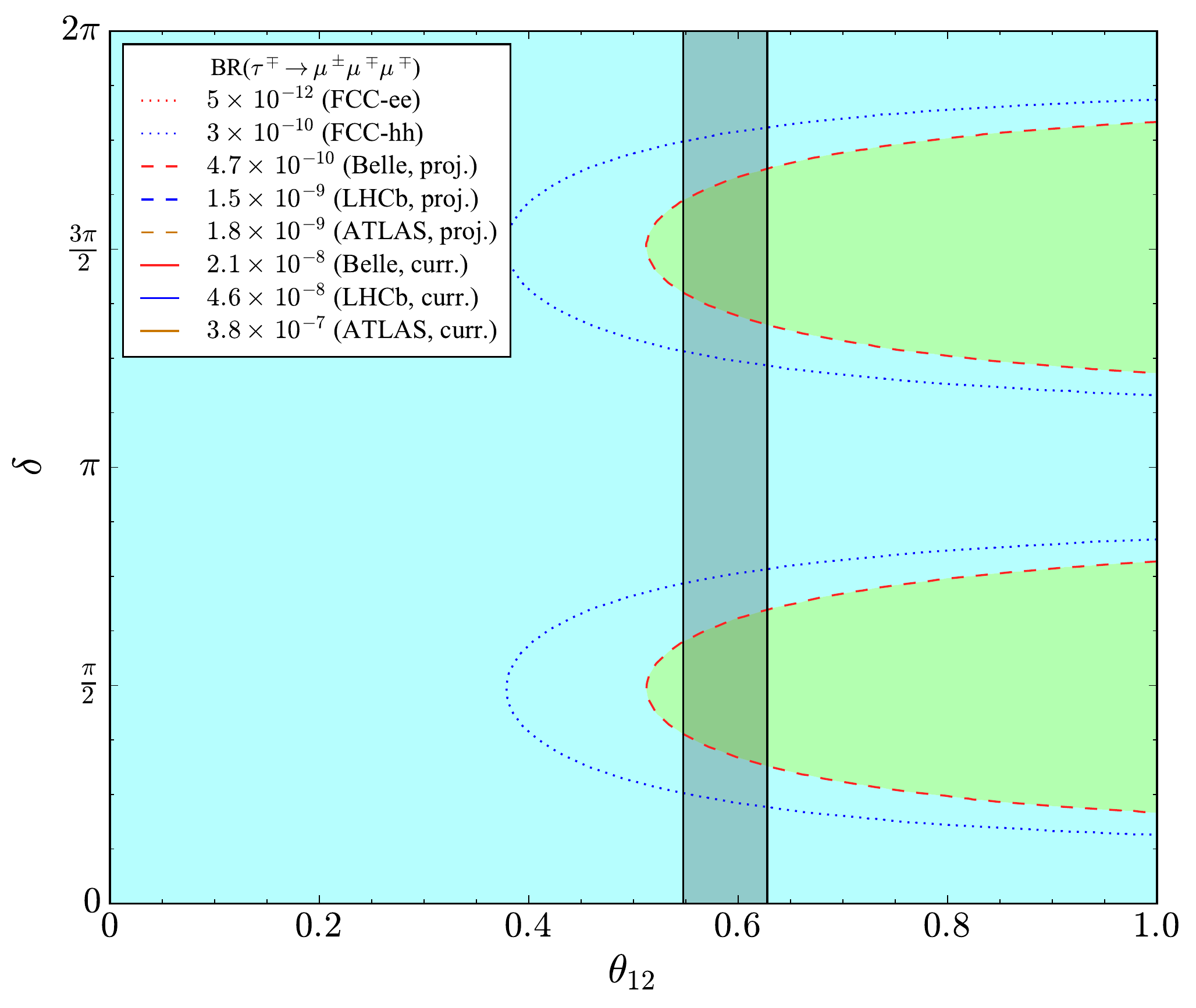}
\caption{}
\label{fig:T2csawneu-tau_3mu}
\end{subfigure}
\hfill
\begin{subfigure}[b]{0.49\textwidth}
\centering
\includegraphics[width=\textwidth]{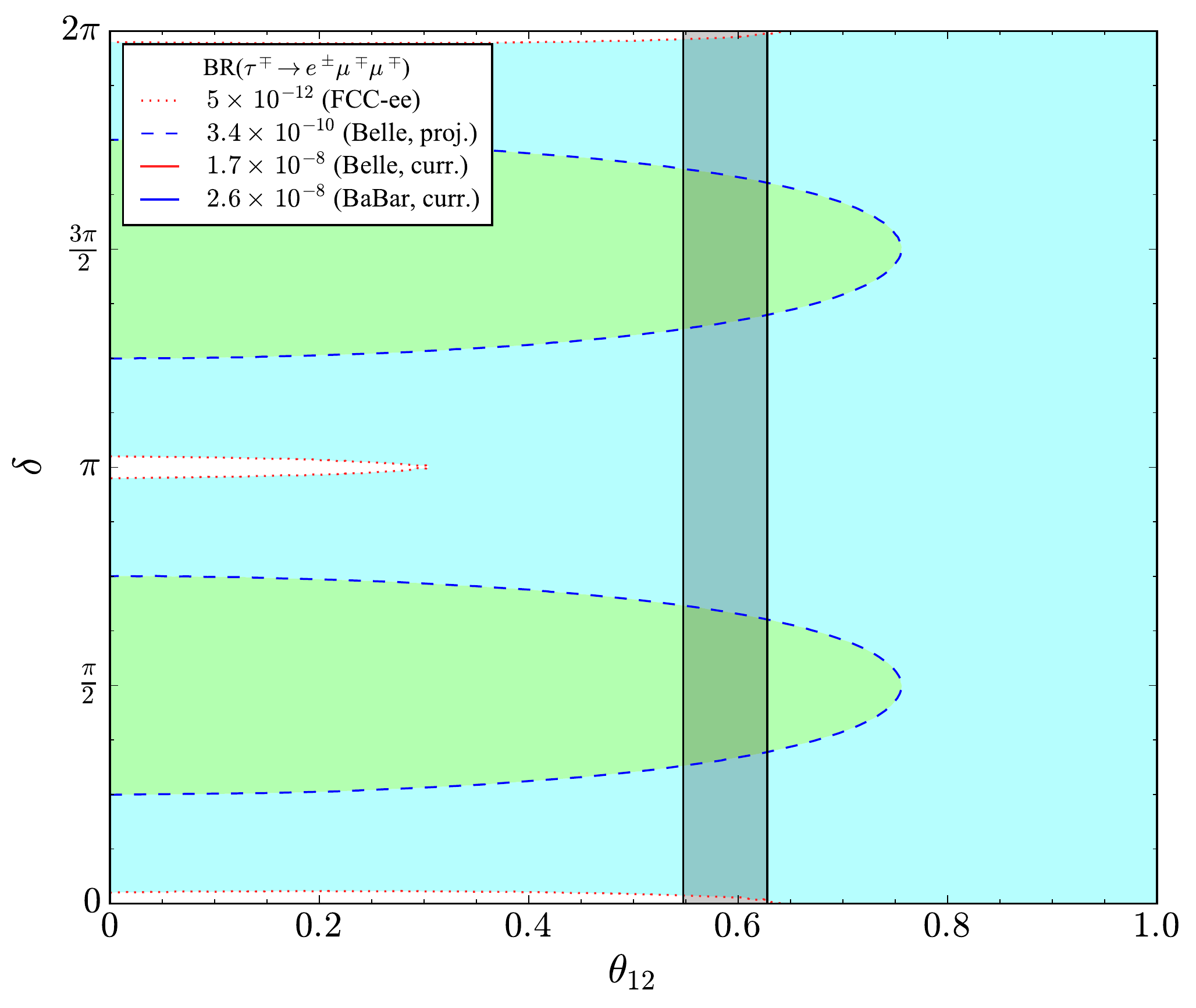}
\caption{}
\label{fig:T2csawneu-tau_e2mu}
\end{subfigure}
\hfill
\caption{Current and future branching ratio limits in the parameter plane of the neutrino oscillation parameter $\theta_{12}$ and the CP violating phase $\delta$ for the Type-II Seesaw Model. (a) Shows the limits from the decay $\tau^{\mp} \to \mu^{\pm} \mu^{\mp} \mu^{\mp}$, and (b) shows the limits from the decay $\tau^\mp \to e^\pm \mu^\mp \mu^\mp$. The dark shaded bands represent the allowed $3\sigma$ values of $\theta_{12}$.}
\label{fig:T2csaw_two}
\end{figure}

In Fig.~\ref{fig:T2csaw_one}, we choose a small $v_{\Delta}$ range, $(10^{-11}-10^{-9})$ GeV, that can naturally explain the small neutrino masses $m_{\nu} \sim (0.01-0.1)$ eV, with $\mathcal{O}(1)$ coupling. For a moderate $v_{\Delta} =10^{-10}$ GeV, and with the neutrino mass $m_{\nu} \sim 0.1$ eV, the present constraints on $\mu_{\Delta}$ and the doubly charged Higgs mass coming from Belle are $\mu_{\Delta} \geq 7.8\times 10^{-9}$ GeV and $m_{\Delta^{\pm \pm}} \geq 1.8$ TeV, using the $\tau^{\mp} \to \mu^{\pm} \mu^{\mp} \mu^{\mp}$ decay. The future experiments Belle-II and FCC-ee could constrain the doubly charged Higgs mass up to $m_{\Delta^{\pm \pm}} \geq 4.6$ TeV and $14.5$ TeV with $\mu_{\Delta} \geq 5.0\times 10^{-8}$ GeV and $4.9\times 10^{-7}$ GeV, respectively. 

The neutrino mass matrix $M_{\nu}$ is diagonalised by the PMNS mixing matrix,
\be
\label{eq:diagonalisation}
U^{*}_{{\mathrm{P}}}M_{\nu}U^{\dagger}_{\mathrm{P}}=M_d~,
\ee
where $M_d$ is the diagonal neutrino mass matrix $M_d=\mathrm{diag}(m_1, m_2, m_3)$, and the PMNS mixing matrix $U_{\mathrm{P}}$ has the following form:
\be
 U_{\mathrm{P}} \!= \! \left(
 \begin{array}{ccc}
 c_{12} \, c_{13} & s_{12}\, c_{13} & s_{13}\, e^{-i \delta}\\
 -c_{23}\, s_{12}-s_{23}\, s_{13}\, c_{12}\, e^{i \delta} &
 c_{23}\, c_{12}-s_{23}\, s_{13}\, s_{12}\, 
e^{i \delta} & s_{23}\, c_{13}\\
 s_{23}\, s_{12}-\, c_{23}\, s_{13}\, c_{12}\, e^{i \delta} &
 -s_{23}\, c_{12}-c_{23}\, s_{13}\, s_{12}\, 
e^{i \delta} & c_{23}\, c_{13}
 \end{array}
 \right) 
\!\! \left(
\begin{array}{ccc}
1 & 0 & 0 \cr
0 & e^{i \alpha_1} & 0 \cr
0 & 0 &  e^{i \alpha_2 }
 \end{array}
\!\!\! \right).~~~
\label{eq:upmns}
\ee
In the above, $s_{ij} \equiv \sin\theta_{ij}$ and $c_{ij} \equiv \cos\theta_{ij}$, where $\theta_{ij}$ are the neutrino oscillation parameters. Furthermore, $\delta$ is the Dirac CP violating phase and $\alpha_{1,2}$ are the Majorana phases. In Fig.~\ref{fig:T2csaw_two}, we allow for a non-zero PMNS phase $\delta$ in the range $0-2\pi$, and investigate the effect of varying $\delta$ along with the neutrino oscillation parameter $\theta_{12}$ on the two decay processes, while fixing the other oscillation parameters to their best-fit values and the lightest neutrino mass to $m_1=0.1$ eV. 
The dark vertical shaded bands show the region of the parameter space allowed by the current $3\sigma$ limits on $\theta_{12}$. For the $\tau^{\mp} \to \mu^{\pm} \mu^{\mp} \mu^{\mp}$ decay,    we consider  $\mu_{\Delta} = 1.5\times 10^{-7}~\mathrm{GeV}$ and $v_{\Delta} = 10^{-10}$ GeV, resulting in  $m_{\Delta^{\pm \pm}}=8.0$ TeV. In the case of  $\tau^\mp \to e^\pm \mu^\mp \mu^\mp$, we use an increased  $\mu_{\Delta} = 2.5\times 10^{-7}~\mathrm{GeV}$ and $v_{\Delta} = 10^{-10}$ GeV, giving  $m_{\Delta^{\pm \pm}}=10.3$ TeV.   
The Belle-II experiment could rule out  $\delta$ in the ranges  $1.1-2.0$ and $4.2-5.1$, while experiments at the FCC-ee 
could exclude all values of $\delta$ for these choices of $\mu_{\Delta}$ and $v_{\Delta}$.  We find similar constraints 
when using the $\theta_{23}-\delta$ contours instead, which we do not show here. 

We conclude this subsection by justifying our approach of only considering limits from the LFV $\tau$ 
lepton decays. The current bound on the branching fraction for $\mu^{\mp} \to e^{\pm} e^{\mp} e^{\mp}$ is 
$\textrm{BR}(\mu^{\mp} \to e^{\pm} e^{\mp} e^{\mp}) \leq 10^{-12}$ \cite{Bellgardt:1987du}. 
This tight bound from $\mu^{\mp} \to e^{\pm} e^{\mp} e^{\mp}$ 
imposes stronger limits in the plane of $\mu_{\Delta}$ and $v_{\Delta}$ than those arising from the 
$\tau$ lepton decays, shown in Fig.~\ref{fig:T2csaw_one}.  However, when varying the neutrino oscillation 
parameters and phases within experimental bounds, it is possible to suppress the branching fraction of 
$\mu^{\mp} \to e^{\pm} e^{\mp} e^{\mp}$ while leaving that of $\tau \to \ell\mu\mu$ essentially unchanged.  
We can consider the oscillation effects by defining the ratio,
\begin{equation}
\mathcal{R}=\frac{\rm{BR(\tau^{\mp} \to \mu^{\pm} \mu^{\mp} \mu^{\mp})}}{\rm{BR(\mu^{\mp} \to e^{\pm} e^{\mp} e^{\mp})}} \propto \frac{|M_{\nu}(\mu, \tau) M_{\nu}(\mu, \mu)|^2}{|M_{\nu}(\mu, e) M_{\nu}(e,e)|^2}~,
\label{r}
\end{equation}
and varying all the oscillation parameters and phases within their allowed $3\sigma$ ranges. For 
quasi-degenerate neutrino masses with an inverted hierarchy spectrum, and with $m_{3}=0.1$ eV, we find 
that $\mathcal{R}$ can be as large as $10^{6}$, due to cancellations in the neutrino mass matrix $M_{\nu}$, 
which is calculated via Eq.~\eqref{eq:diagonalisation}. Such regions of the parameter space suppress the 
branching ratio of $\mu^{\mp} \to e^{\pm} e^{\mp} e^{\mp}$ enough so that the strongest limits on $\mu_{\Delta}$ 
and $v_{\Delta}$ arise from the LFV $\tau$ lepton decays, which can remain largely unaffected. Therefore, 
Fig.~\ref{fig:T2csaw_one} qualitatively demonstrates the constraints that can be obtained in regions where 
the LFV $\tau$ lepton decays provide the dominant source of all LFV decays. 

\subsection{Left-Right Symmetric Model}

The minimal Left-Right Symmetric Model is based on the  gauge group $SU(3)_c\times SU(2)_L\times SU(2)_R\times U(1)_{B-L} $ \cite{Pati:1974yy, Mohapatra:1974gc, Senjanovic:1975rk, Duka:1999uc}. The fermions are assigned in the doublet representations of $SU(2)_L$ and $SU(2)_R$. In addition to the particle content of the Standard Model, the model contains three right-handed Majorana neutrinos {$N_R$} paired with the charged leptons $l_R$, and the additional gauge bosons $W_R$ and $Z^{\prime}$. The Higgs fields correspond to  a bi-doublet $\Phi$ and two Higgs triplets $\Delta_L$ and $\Delta_R$ with the following quantum numbers under the gauge group:  $\Phi (1, 2, 2, 0)$,  $\Delta_L (1, 3, 1, 2)$ and $\Delta_R (1, 1, 3, 2)$. The Higgs triplet $\Delta_R$ takes the vacuum expectation value $v_R$ and spontaneously breaks $SU(2)_R\times U(1)_{B-L}$  down to the group $U(1)_Y$ of the SM. This generates the masses of the $W_R$ and $Z^{\prime}$ gauge bosons and the masses of the right-handed neutrinos. The neutral components of the bi-doublet field $\Phi$ also acquire a vev, which is denoted as $\langle\Phi\rangle=\mathrm{diag}(\kappa_1, \kappa_2)/\sqrt{2}$, and this breaks the electroweak symmetry down to $U(1)_Q$, giving masses to the quarks and leptons. 

The Higgs triplet $\Delta_R$ couples to the right-handed neutrinos $N_R$ and  generates the Majorana masses of the heavy neutrinos during the symmetry breaking. The light neutrino 
masses are generated as a sum of two seesaw contributions, one suppressed by the right-handed 
neutrino mass (Type-I) \cite{Minkowski:1977sc, Mohapatra:1979ia, Yanagida:1979as, GellMann:1980vs, Schechter:1980gr} and the other suppressed by the Higgs triplet mass (Type-II) \cite{Magg:1980ut, Lazarides:1980nt}. The different vevs of the bi-doublets and triplets follow the hierarchy $v_L\ll \kappa_{1,2} \ll v_R$. Below, we discuss the different neutrino masses and the Higgs sector of the LRSM in detail, and their contribution to the tree-level LFV processes $\tau^{\mp} \to \mu^{\pm} \mu^{\mp} \mu^{\mp}$ and $\tau^\mp \to e^\pm \mu^\mp \mu^\mp$. 

\subsubsection{Neutrino mass}  
The Yukawa Lagrangian in the lepton sector has the following form:
\begin{eqnarray}
-{\cal L}_Y &=& h\bar{\psi}_{L}\Phi \psi_{R} 
+ \tilde{h} \bar{\psi}_{L}\tilde{\Phi} \psi_{R}
+ f_{L} \psi_{L}^{\mathrm{T}} C i\tau_2 \Delta_L \psi_{L} \nonumber \\ 
&& + f_{R} \psi_{R}^{\mathrm{T}} C i\tau_2 \Delta_R \psi_{R} 
+{\rm h.c.}~,
\label{eq:yuk}
\end{eqnarray}
where  $C$ is the charge-conjugation matrix,  $C=i\gamma_2\gamma_0$, and $\tilde{\Phi}=\tau_2\Phi^*\tau_2$, with $\tau_2$ being the second Pauli matrix. Furthermore,  $h, \tilde{h}, f_L $ and $f_R$ are the Yukawa couplings. After symmetry breaking, the Yukawa Lagrangian generates the neutrino mass matrix,
\begin{eqnarray}
{\cal M}_\nu =
\begin{pmatrix}
M_L & M_D \\
M_D^{\mathrm{T}} & M_R
\end{pmatrix}~.
\label{eq:big}
\end{eqnarray} In the seesaw approximation, this leads to the following light and heavy neutrino mass matrices  (up to $\mathcal{O}(M^{-1}_R)$) \cite{Grimus:2000vj}:
\begin{eqnarray}
M_\nu  & \approx &  M_L - M_D M_R^{-1} M_D^{ \mathrm{T}} =  \sqrt 2 v_L f_L - \frac{\kappa^2}{\sqrt 2 v_R} h_D f_R^{-1} h_D^{ \mathrm{T}}~,
\label{eq:mnu}
\end{eqnarray}
and 
\begin{eqnarray}
M_R = \sqrt 2 v_R f_R ~,
\end{eqnarray}
where $\kappa= \sqrt{\kappa^2_1+\kappa^2_2}$, $M_L = \sqrt 2 v_L f_L$ and the Dirac mass is $M_D = h_D \kappa=\left(\kappa_1 h + \kappa_2 \tilde{h} \right)/\sqrt{2}$. The mass matrix given in Eq.~\eqref{eq:big} can be diagonalised 
by a $6\times 6$ unitary matrix as follows: 
\be
{\cal V}^{\mathrm{T}}{\cal M}_\nu {\cal V} \ = \ \left(\begin{array}{cc} \widetilde{M}_\nu & {\bf 0} \\ {\bf 0} & \widetilde{M}_R \end{array}\right)~,
\ee
where ${\widetilde{M}_\nu}={\rm{diag}}(m_1,m_2,m_3)$ and $\widetilde{M}_R = {\rm diag}(m_{N_4},m_{N_5},m_{N_6})$.  In the subsequent analysis, we denote the mixing matrix as,
\be
\cal{V} =
\begin{pmatrix}
 U & S \\
T & V 
\end{pmatrix}~.
\label{mat}
\ee
The Yukawa interaction of the doubly charged Higgs with the two charged leptons that mediates the LFV processes $\tau^{\mp} \to \mu^{\pm} \mu^{\mp} \mu^{\mp} $ and $\tau^\mp \to e^\pm \mu^\mp \mu^\mp$ is given by, 
\be
\mathcal{L}_Y=f_L \bar{l}_L^c \delta^{++}_L l_L + f_R \bar{l}^c_R \delta^{++}_R l_R~+\rm{h.c}.~.
\label{yukawalr}
\ee
We note that imposing the discrete parity or charge conjugation as a symmetry along with $SU(2)_R\times U(1)_{B-L}$ will lead to $f_L = f_R$ or $f_L=f^*_R$,  and a hermitian or symmetric $M_D$, respectively. As we will show in the next subsection, among the two Higgs triplets $\delta^{\pm \pm}_L $ and $\delta^{\pm \pm}_R$, the right-handed triplet gives the dominant contribution to the tree-level flavour violating processes due to our choice of Higgs masses. Hence, the dominant contribution in the Lagrangian can be approximated as, 

\be
\mathcal{L}_Y \approx  \frac{M_R}{\sqrt{2}v_R} \bar{l}^c_R \delta^{++}_R l_R = \frac{V^*_R \widetilde{M}_R V^{\dagger}_R }{\sqrt{2}v_R} \bar{l}^c_R \delta^{++}_R l_R~+\rm{h.c.}~,
\label{yukawalr2}
\ee
where $V_R$ is the diagonalising matrix for the heavy neutrino mass matrix $M_R$, $V^T_R M_R V_R = \widetilde{M}_R$, and $V \sim V_R$ \cite{Grimus:2000vj}. A detailed discussion on LFV for this model for all other modes can be found in \cite{Bonilla:2016fqd,Barry:2013xxa}.

\subsubsection{Higgs mass}

We now discuss the scalar potential and Higgs spectrum in detail. The LRSM consists of the two scalar triplets and one bi-doublet field, that after left-right and electroweak symmetry breaking  leads to  fourteen physical Higgs states. Among them, a few of the neutral Higgs bosons are required to be heavier than several tens of TeV and do not contribute to the tree-level LFV processes. We follow a simplified approach by judiciously choosing the parameter space, where the doubly charged Higgs arising from $\Delta_R$  is lighter than the other BSM Higgs states, and hence gives the dominant contribution in the tree-level LFV processes. 

The  scalar potential for the LRSM has the following form \cite{Deshpande:1990ip, Roitgrund:2014zka, Maiezza:2016ybz}: 

\begin{eqnarray}
 V(\Phi,\Delta_L,\Delta_R) &=& -\mu_1^2 \Tr{\Phi^\dagger\Phi}
 - \mu_2^2\Tr{\Phi^\dagger\tilde{\Phi} + \tilde{\Phi}^\dagger\Phi}
 - \mu_3^2\Tr{\Delta_L^\dagger\Delta_L + \Delta_R^\dagger\Delta_R} \nonumber\\ 
 &&  + \lambda_1\left[\Tr{\Phi^\dagger\Phi}\right]^2 +
   \lambda_2\left[\Tr{\Phi^\dagger\tilde{\Phi}}\right]^2 + \lambda_2\left[\Tr{\tilde{\Phi}^\dagger\Phi}\right]^2 
 \nonumber\\   && +
  \lambda_3\Tr{\Phi^\dagger\tilde{\Phi}}\Tr{\tilde{\Phi}^\dagger\Phi} + 
  \lambda_4\Tr{\Phi^\dagger\Phi}\Tr{\Phi^\dagger\tilde{\Phi} + \tilde{\Phi}^\dagger\Phi} \nonumber\\
 && +\rho_1\left[\Tr{\Delta_L^\dagger\Delta_L}\right]^2 + 
 \rho_1\left[\Tr{\Delta_R^\dagger\Delta_R}\right]^2 +
 \rho_3\Tr{\Delta_L^\dagger\Delta_L}\Tr{\Delta_R^\dagger\Delta_R}
  \nonumber\\  && + 
 \rho_2\Tr{\Delta_L\Delta_L}\Tr{\Delta_L^\dagger\Delta_L^\dagger} + \rho_2\Tr{\Delta_R\Delta_R}\Tr{\Delta_R^\dagger\Delta_R^\dagger}
  \nonumber\\ && +
 \rho_4\Tr{\Delta_L\Delta_L}\Tr{\Delta_R^\dagger\Delta_R^\dagger} + \rho_4\Tr{\Delta_L^\dagger\Delta_L^\dagger}\Tr{\Delta_R\Delta_R} \nonumber\\
&&  \alpha_1\Tr{\Phi^\dagger\Phi}\Tr{\Delta_L^\dagger\Delta_L + \Delta_R^\dagger\Delta_R} +
 \alpha_3 \Tr{\Phi\Phi^\dagger\Delta_L\Delta_L^\dagger + \Phi^\dagger\Phi\Delta_R\Delta_R^\dagger} 
   \nonumber\\ && +
 \left\{
 \alpha_2 e^{i\delta_2}\Tr{\Phi^\dagger\tilde{\Phi}}\Tr{\Delta_L^\dagger\Delta_L} + 
 \alpha_2 e^{i\delta_2}\Tr{\tilde{\Phi}^\dagger\Phi}\Tr{\Delta_R^\dagger\Delta_R} + \text{h.c.}\right\} \nonumber\\ 
&&  + \beta_1\Tr{\Phi^\dagger\Delta_L^\dagger\Phi\Delta_R + \Delta_R^\dagger\Phi^\dagger\Delta_L\Phi} +
 \beta_2\Tr{\Phi^\dagger\Delta_L^\dagger\tilde{\Phi}\Delta_R + \Delta_R^\dagger\tilde{\Phi}^\dagger\Delta_L\Phi}
 \nonumber\\  && +
 \beta_3\Tr{\tilde{\Phi}^\dagger\Delta_L^\dagger\Phi\Delta_R + \Delta_R^\dagger\Phi^\dagger\Delta_L\tilde{\Phi}}.
\end{eqnarray}
The model contains 14 physical Higgs states denoted as $h$, $H^0_{1,2,3}$, $A^{0}_{1,2}$, $H^{\pm}_1$, $H^{\pm}_2$, $\delta^{\pm \pm}_L$, and $\delta^{\pm \pm}_R$ with the masses,  
\begin{eqnarray} 
 m^2_h \approx (125\GeV)^2 &\approx& 2\kp^2\left(\lambda_1 + 4\frac{\kappa_1^2\kappa_2^2}{\kp^4}(2\lambda_2+\lambda_3) + 4\lambda_4\frac{\kappa_1\kappa_2}{\kp^2}\right)~,
 \nonumber\\
M^2_{H^0_1}=M^2_{A^0_1}\approx\alpha_3 \frac{v^2_R}{2} \frac{\kp^2}{\km^2}~,	&\quad& 
M^2_{H^0_3}=M^2_{A^0_2}\approx(\rho_3-2\rho_1) \frac{v^2_R}{2}~,			 \quad 
M^2_{H^0_2}\approx2 \rho_1 v^2_R~,
\nonumber\\
M^2_{H^{\pm}_1}\approx(\rho_3-2\rho_1) \frac{v^2_R}{2} + \alpha_3 \frac{\km^2}{4}~,	&\quad& 
M^2_{\delta^{\pm \pm}_L}\approx(\rho_3-2\rho_1) \frac{v^2_R}{2} + \alpha_3 \frac{\km^2}{2}~,
\nonumber\\
M^2_{H^{\pm}_2}\approx\alpha_3 \frac{v^2_R}{2} \frac{\kp^2}{\km^2}+\alpha_3 \frac{\km^2}{4}~,	&\quad&
M^2_{\delta^{\pm \pm}_R}\approx2 \rho_2v^2_R+\alpha_3 \frac{\km^2}{2}~.
\label{eq:higgsMasses}
\end{eqnarray}
We note that the scalar states $H^0_1$ and $H^0_3$  interact with both the up and down quark sectors and hence mediate the $\Delta F=2$ flavour transitions in the neutral $K$ and $B$ mesons \cite{Zhang:2007da, Maiezza:2014ala, Bertolini:2014sua, Maiezza:2016bzp}. To avoid the flavour-changing neutral Higgs (FCNH) constraints, the neutral Higgs states $H^0_1$, $H^0_3$ and $A^0_{1,2}$ are required to be heavier than  20 TeV \cite{Zhang:2007da, Maiezza:2014ala, Bertolini:2014sua, Maiezza:2016bzp}. We also consider the other neutral Higgs state $H^0_2$  to be heavy in order to be in agreement with the heavy Higgs searches at the LHC. In the Higgs spectrum, we consider  the case where the right-handed doubly charged Higgs boson  is somewhat lighter than the other BSM Higgs states and hence significantly contributes to the LFV processes. We consider the following two benchmark scenarios, BP1 and BP2, with a lower and a higher symmetry breaking scale $v_R$ respectively: 

\begin{itemize}
\item
BP1: $\qquad \alpha_3 = 18.88~, \qquad v_R=8.68~\mathrm{TeV}~,$


\item

BP2: $\qquad \alpha_3 = 1.00~, \qquad v_R=30.00~\mathrm{TeV}~.$

\end{itemize}
For both of the benchmark scenarios, we consider the right-handed mixing matrix $V_R$ to be non-diagonal with unit entries everywhere.  In order for $v_R$ to be less than 10 TeV, the FCNH constraints on the neutral Higgs bosons necessarily require  $\alpha_3$ to be large ($\alpha_3 \sim 8$). Conversely, when $\alpha_3$ is well within the perturbative limit, the FCNH constraints on the neutral Higgs bosons demand a large value of the symmetry breaking scale $v_R$ \cite{Maiezza:2016bzp}. In our analysis we consider the two possibilities, both the large and the natural  $\alpha_3$, and show the restrictions that can be obtained on the heavy neutrino masses and the $\rho_2$ parameter.


\subsubsection{Limits from the LFV branching ratios}

\begin{figure}[!tb]
\centering
\begin{subfigure}[b]{0.49\textwidth}
\centering
\includegraphics[width=\textwidth]{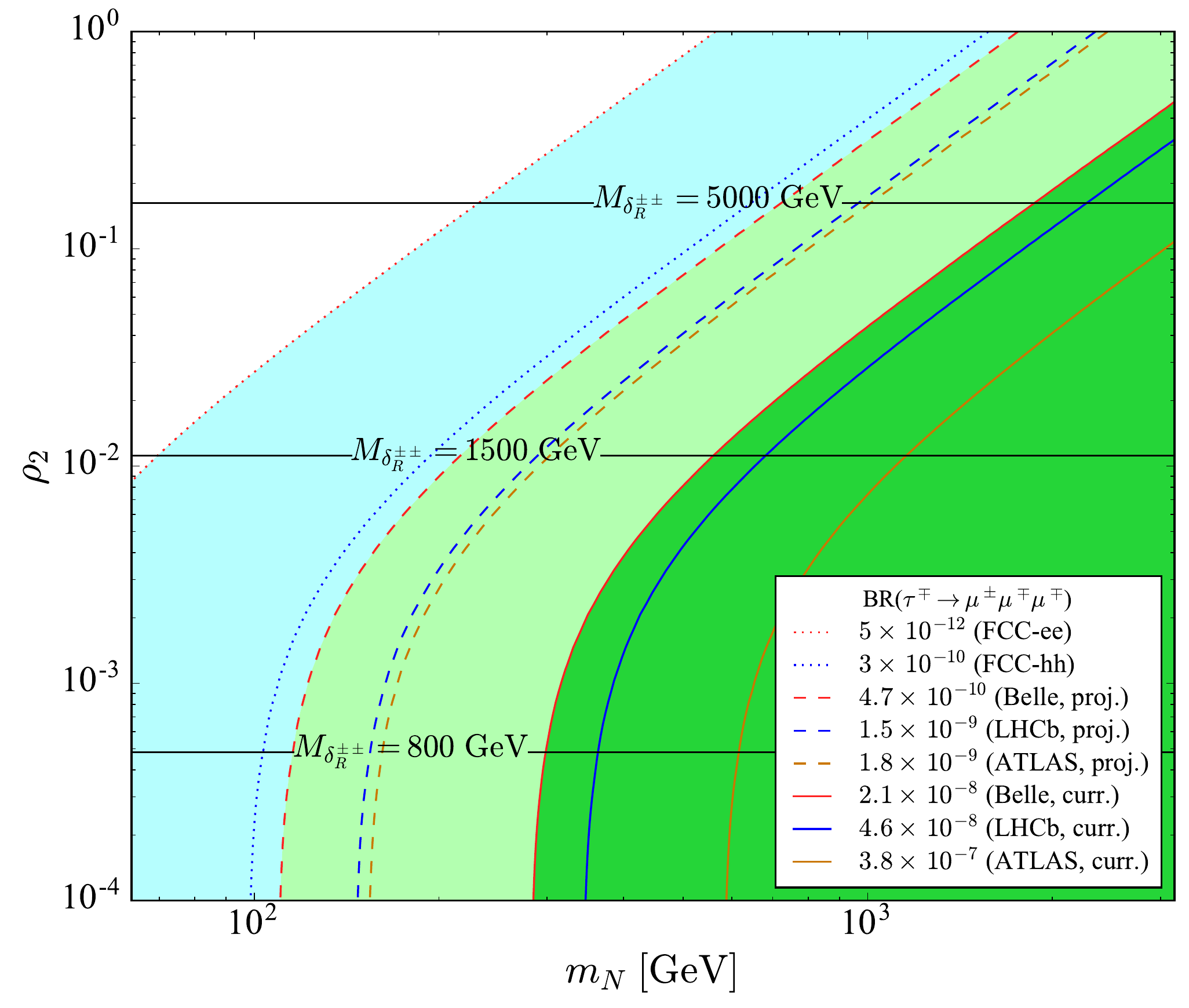}
\caption{}
\label{fig:MLRSM-tau_3mu}
\end{subfigure}
\hfill
\begin{subfigure}[b]{0.49\textwidth}
\centering
\includegraphics[width=\textwidth]{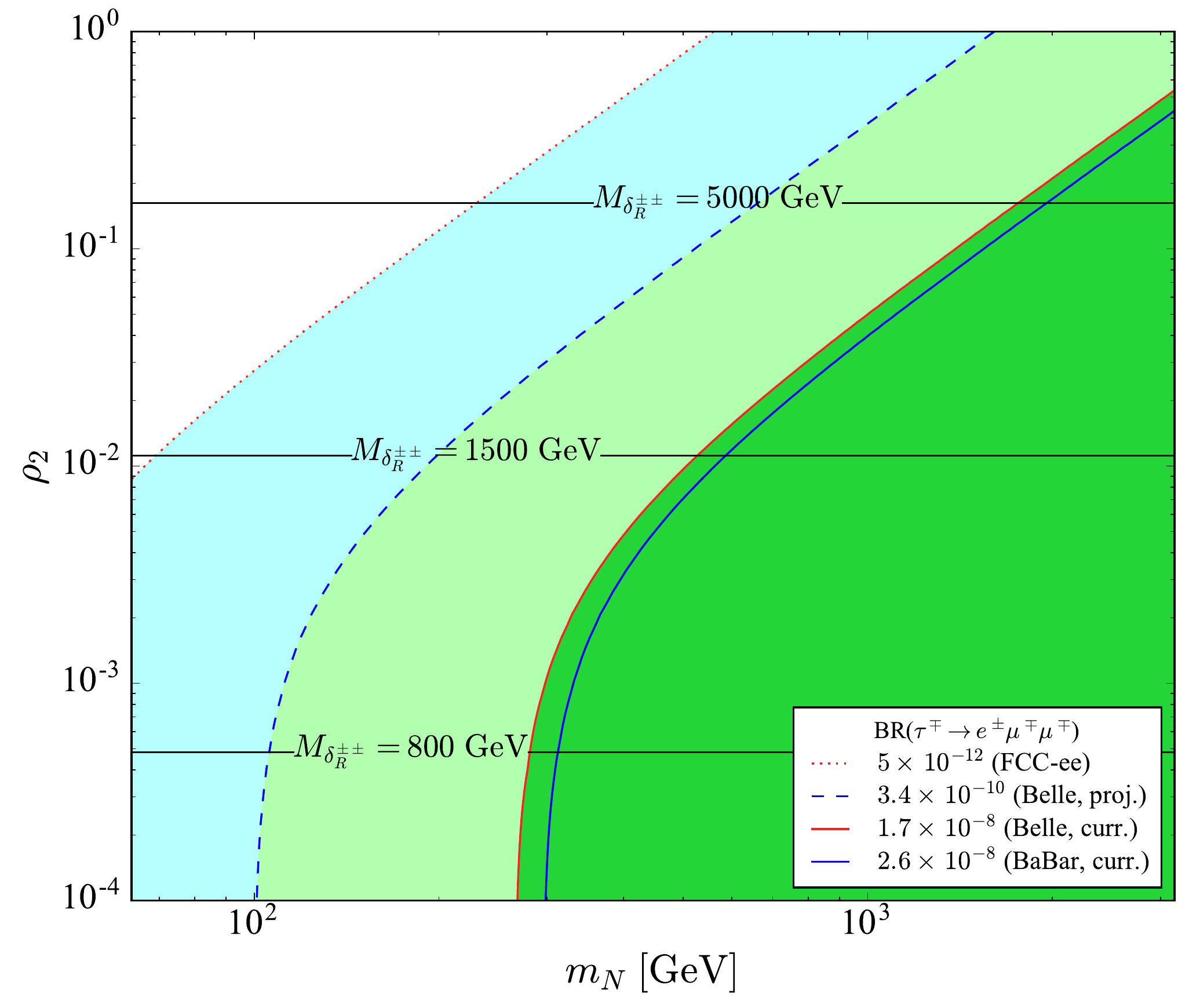}
\caption{}
\label{fig:MLRSM-tau_e2mu}
\end{subfigure}
\hfill
\caption{Current and future branching ratio limits in the parameter plane of the right-handed neutrino masses $m_N$ and the parameter $\rho_2$ for the LRSM for the benchmark scenario BP1. (a) Shows the limits from the decay $\tau^{\mp} \to \mu^{\pm} \mu^{\mp} \mu^{\mp}$, and (b) shows the limits from the decay $\tau^\mp \to e^\pm \mu^\mp \mu^\mp$. The solid black lines represent constant values of the mass of the doubly charged Higgs $\delta^{\pm \pm}_R$.}
\label{fig:MLRSM_one}
\end{figure}

\begin{figure}[!tb]
\centering
\begin{subfigure}[b]{0.49\textwidth}
\centering
\includegraphics[width=\textwidth]{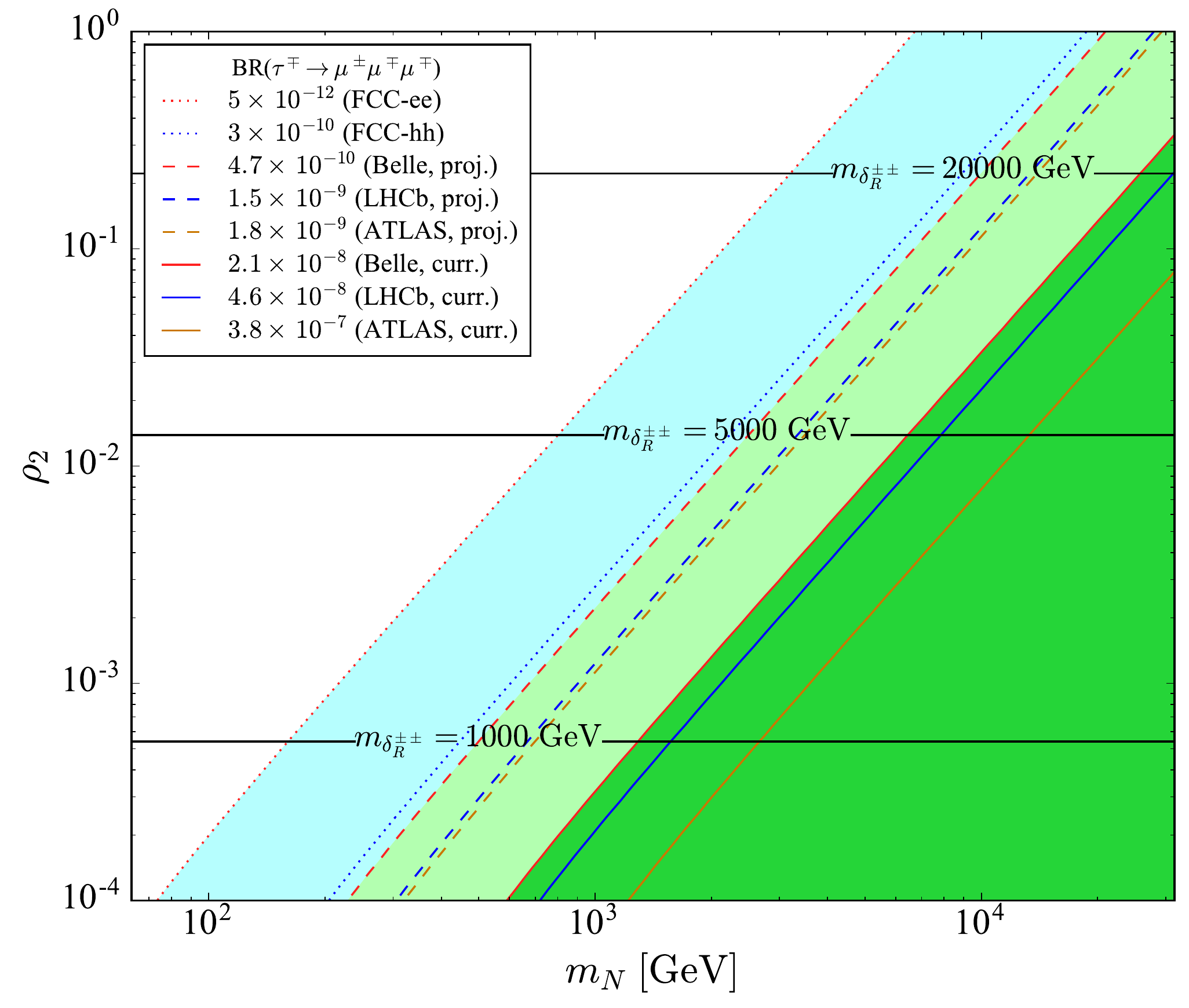}
\caption{}
\label{fig:MLRSMlowalpha3-tau_3mu}
\end{subfigure}
\hfill
\begin{subfigure}[b]{0.49\textwidth}
\centering
\includegraphics[width=\textwidth]{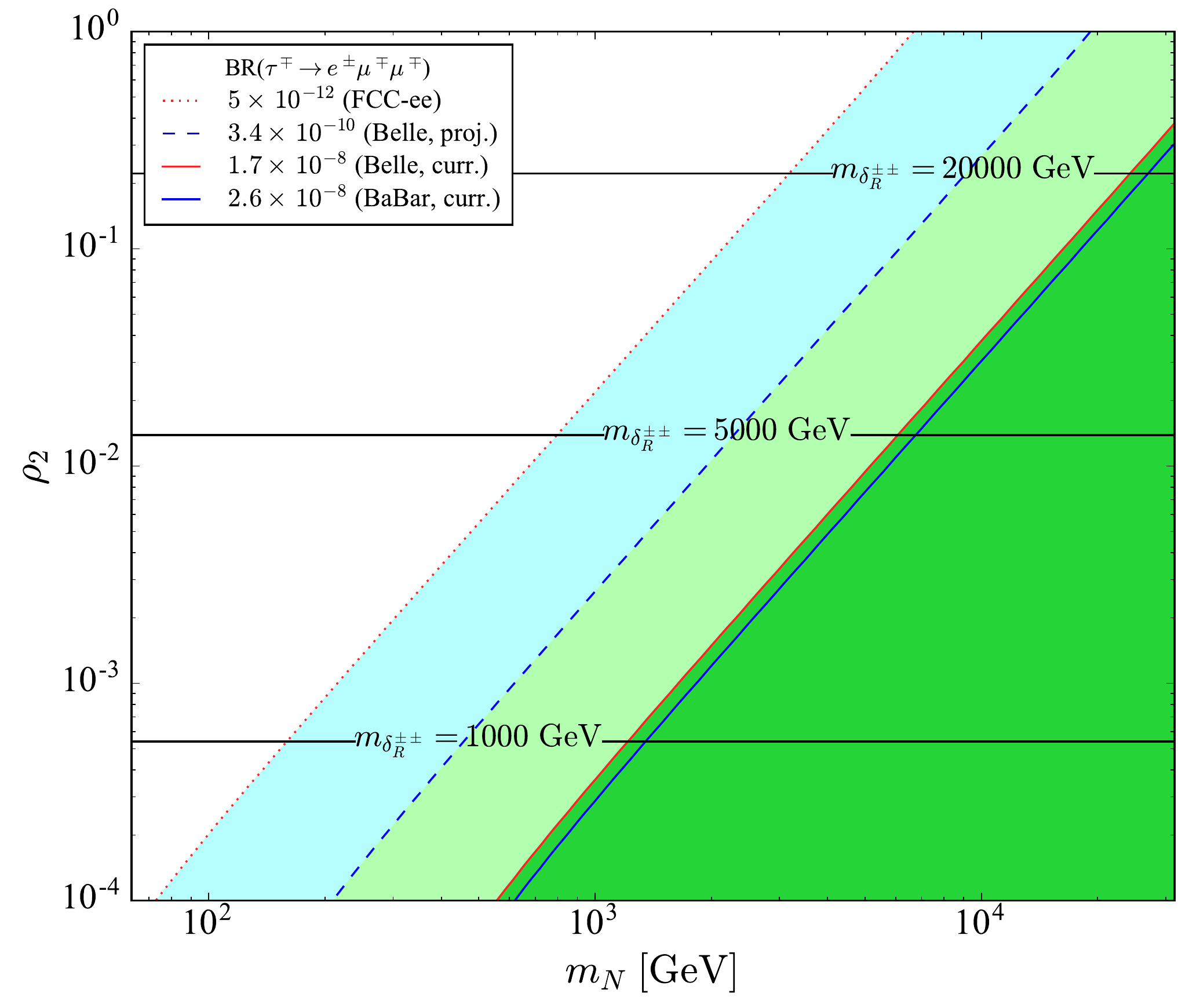}
\caption{}
\label{fig:MLRSMlowalpha3-tau_e2mu}
\end{subfigure}
\hfill
\caption{Current and future branching ratio limits in the parameter plane of the right-handed neutrino masses $m_N$ and the parameter $\rho_2$ for the LRSM for the benchmark scenario BP2. (a) Shows the limits from the decay $\tau^{\mp} \to \mu^{\pm} \mu^{\mp} \mu^{\mp}$, and (b) shows the limits from the decay $\tau^\mp \to e^\pm \mu^\mp \mu^\mp$. The solid black lines represent constant values of the mass of the doubly charged Higgs $\delta^{\pm \pm}_R$.}
\label{fig:MLRSM_two}
\end{figure}

The two doubly charged Higgs states $\delta^{\pm \pm}_L$ and $\delta^{\pm \pm}_R$ mediate the $\tau \to l_i l_j l_k$  process at tree-level. The amplitude for the LFV process $\tau^{\mp} \to \mu^{\pm} \mu^{\mp} \mu^{\mp}$  is proportional to the coefficient $C_{\tau \mu \mu \mu}$, which is defined as,
\begin{eqnarray}
C_{\tau  \mu \mu \mu} &=& \frac{{f_L}_{ \tau \mu} {f_L}_{\mu \mu}}{M^2_{\delta^{\pm \pm }_L}}+\frac{{f_R}_{\tau \mu} {f_R}_{\mu \mu}}{M^2_{\delta^{\pm \pm}_R}}~.
\label{pwdtau3mut}
\end{eqnarray}
Since in our case the chosen parameter  $M_{\delta^{\pm \pm}_L}$ is much heavier than $M_{\delta^{\pm \pm}_R}$, the dominant contribution arises due to $\delta^{\pm \pm}_R$, 
\begin{eqnarray}
C_{\tau  \mu\mu\mu} &=& \frac{{f_R}_{\tau \mu} {f_R}_{\mu \mu}}{M^2_{\delta^{\pm \pm}_R}} \approx \frac{{M_R}_{\tau \mu} {M_R}_{\mu \mu}}{2 v^2_R M^2_{\delta^{\pm \pm}_R}}~=\frac{({V^*_R\widetilde{M}_R V^{\dagger}_R})_{\tau \mu}({V^*_R\widetilde{M}_R V^{\dagger}_R})_{\mu \mu}}{2 v^2_R (2 \rho_2v^2_R+\alpha_3 \frac{k^2_{-}}{2})}~.
\label{pwdtau3mur}
\end{eqnarray}
The amplitude for the LFV process  $\tau^\mp \to e^\pm \mu^\mp \mu^\mp$ can be obtained by replacing the ${ \tau \mu}$ element in Eq.~\eqref{pwdtau3mur} with the $ \tau e $ element. A limit on the branching ratio of the flavour violating decays will constrain the  doubly charged Higgs mass from below and the right-handed neutrino mass from above.  In Fig.~\ref{fig:MLRSM_one}, corresponding to BP1, we show the branching ratio limits for the case where the three right-handed neutrino masses are all equal and denoted by $m_N$, and are varied along with the parameter $\rho_2$. In Fig.~\ref{fig:MLRSM_two}, we show the equivalent plots for BP2.  For BP1, the current limit from Belle imposes the constraint on the right-handed neutrino masses $m_N \leq 290$ GeV for the doubly charged Higgs mass $M_{\delta_R^{\pm \pm}}= 420 $ GeV for the $\tau^{\mp} \to \mu^{\pm} \mu^{\mp} \mu^{\mp}$ and $\tau^{\mp} \to e^{\pm} \mu^{\mp} \mu^{\mp}$ decays. This $M_{\delta^{\pm \pm}_R}$ mass is the lower limit set  by the 13 TeV ATLAS search for the right-handed triplet \cite{ATLAS-CONF-2016-051}. For BP2, with a higher value of the symmetry breaking scale $v_R$, the mass limits are  much higher:  $m_N \lesssim 10 $ TeV for the doubly charged Higgs mass $M_{\delta_R^{\pm \pm}}=8$ TeV. For both of the scenarios, a future circular collider will be able to probe much smaller values of $m_N$.

\begin{figure}[!tb]
\centering
\begin{subfigure}[b]{0.49\textwidth}
\centering
\includegraphics[width=\textwidth]{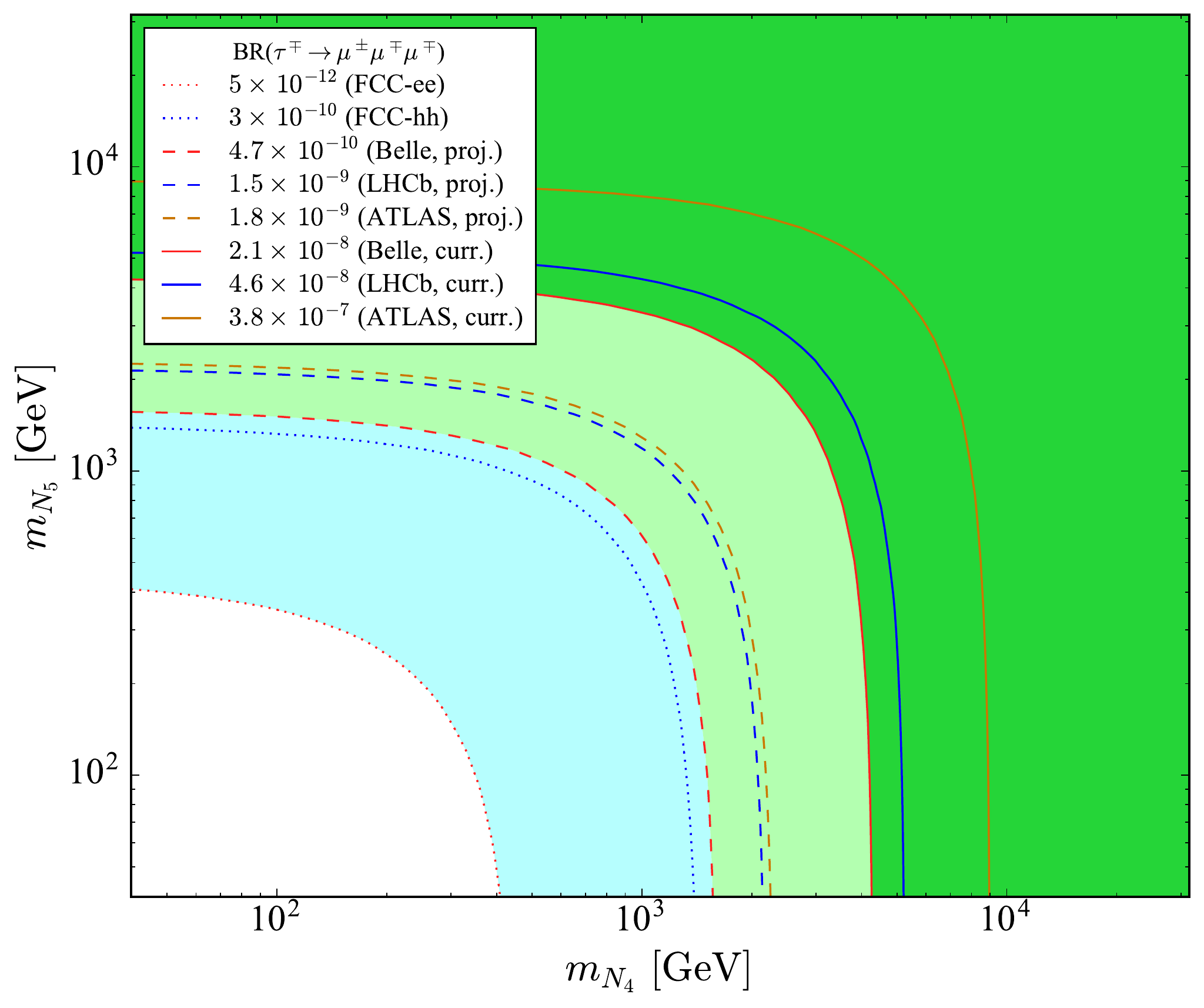}
\caption{}
\label{fig:MLRSMneu-tau_3mu}
\end{subfigure}
\hfill
\begin{subfigure}[b]{0.49\textwidth}
\centering
\includegraphics[width=\textwidth]{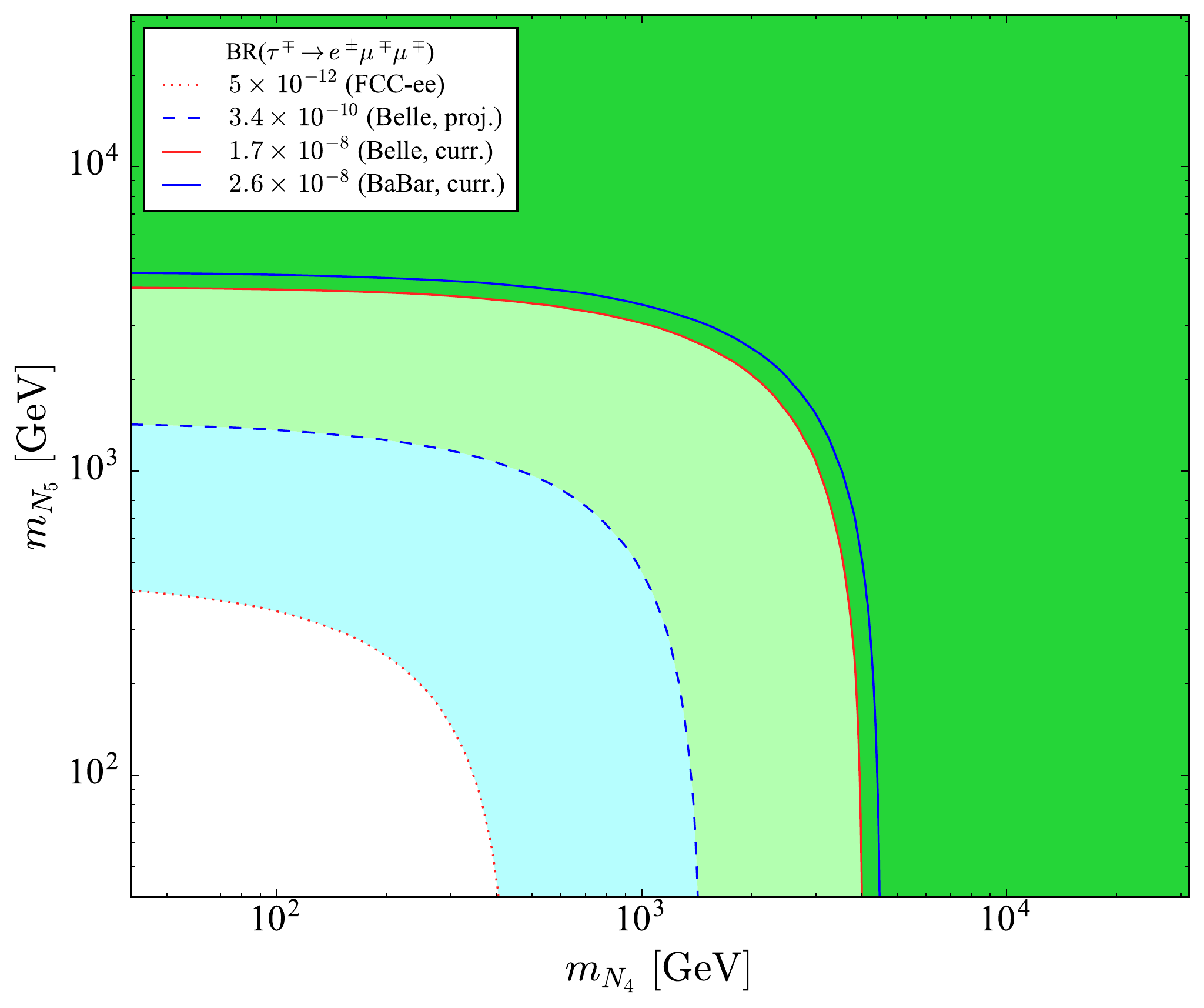}
\caption{}
\label{fig:MLRSMneu-tau_e2mu}
\end{subfigure}
\hfill
\caption{Current and future branching ratio limits in the parameter plane of the right-handed neutrino masses $m_{N_4}$ and $m_{N_5}$ for the LRSM. (a) Shows the limits from the decay $\tau^{\mp} \to \mu^{\pm} \mu^{\mp} \mu^{\mp}$, and (b) shows the limits from the decay $\tau^\mp \to e^\pm \mu^\mp \mu^\mp$.}
\label{fig:MLRSM_three}
\end{figure}

In Fig.~\ref{fig:MLRSM_three}, we consider the scenario of non-degenerate right-handed neutrino masses $m_{N_{4,5,6}}$. We show the branching ratio limits in the plane of the right-handed neutrino masses $m_{N_4}$ and $m_{N_5}$ for the case of BP1, while fixing  $m_{N_6} = 100~\mathrm{GeV}$ and the doubly charged Higgs mass $M_{\delta^{\pm \pm}_R}=4$ TeV. The present stringent limit from Belle constrains both of the $m_{N_4}$ and $m_{N_5}$ masses to be smaller than $\sim$ 1 TeV, while the FCC-ee could probe these masses down to $\sim 100$ GeV.

In our analysis, we considered the possibilities of both a lower and a higher symmetry breaking scale $v_R$. While a lower symmetry breaking scale and a right-handed gauge boson with mass $M_{W_R}$ $\lesssim (5-6)$ TeV is within the reach of the 13 TeV LHC, a higher symmetry breaking scale, such as that in BP2, along with a much heavier $W_R$ could be probed at a 100 TeV future circular collider \cite{Dev:2016dja, Maiezza:2016ybz}. In \cite{Chakrabortty:2016wkl, Maiezza:2016ybz}, the impact of renormalisation group evolution of the quartic couplings on the discovery of $W_R$ and the Higgs states has been discussed and bounds on the quartic couplings have been derived by analysing stability conditions.  A lower symmetry breaking scale with a $W_R$ accessible  at the 13 TeV LHC implies a larger  $\rho_2$ (for a cut-off scale 10$M_{W_R}$ with $M_{W_R}=6$ TeV, then $\rho_2 \ge 0.35$  \cite{Maiezza:2016ybz}) and hence a larger $M_{\delta^{\pm \pm}_R}$. This cannot be directly produced at the LHC, but instead can be tested through indirect detection. Conversely, for a larger symmetry breaking scale with $M_{W_R} \sim (20-30)$ TeV the bounds on $\rho_2$ are relaxed. In our discussion, we do not specify any particular mass of the other Higgs states and the cut-off scale of the theory. Instead, we  independently analyse the implication of the branching ratio limits for the flavour violating processes $\tau^{\mp } \to \mu^{\pm } \mu^{\mp} \mu^{\mp}$ and $\tau^{\mp}\to e^{\pm} \mu^{\mp}\mu^{\mp}$ on the relevant model parameter $\rho_2$ and the doubly charged Higgs mass $M_{\delta^{\pm \pm}_R}$.

\subsection{Minimal Supersymmetric Standard Model}
 
Within the MSSM the soft supersymmetry breaking parameters in the slepton sector are a generic source of lepton flavour violation. Without assuming a specific SUSY breaking mechanism that ensures a suppression of off-diagonal terms in the slepton mass matrix, their presence can induce a misalignment in flavour space between the lepton and slepton mass matrices, which cannot be rotated away. 

The non-diagonal hermitian $6 \times 6$ slepton mass matrix receives contributions from $D$, $F$, $A$ and $M$ terms \cite{Martin:1997ns}, where the latter two can induce mixing between different slepton generations. In the electroweak interaction basis $(\tilde{e}_L,\tilde{\mu}_L,\tilde{\tau}_L,\tilde{e}_R,\tilde{\mu}_R,\tilde{\tau}_R)$, the slepton mass matrix has the following form: 
\begin{equation}
\mathcal{M}^2_{\tilde{l}} = \begin{pmatrix} M^2_{\tilde{l}\,LL} & M^2_{\tilde{l}\,LR} \\ M^{2\,\dagger}_{\tilde{l}\,LR} & M^2_{\tilde{l}\,RR}
\end{pmatrix}~,
\end{equation}
where  each of the $M^2_{\tilde{l}\,LL}$, $M^2_{\tilde{l}\,RR}$, $M^2_{\tilde{l}\,LR}$ and $M^2_{\tilde{l}\,RL}$ is a $3\times 3$ matrix, i.e. 
\begin{align} 
M^2_{\tilde{l}\,LL\,ij} &= m^2_{\tilde{L}\,ij} + \left(m^2_{l_i}+(-\frac{1}{2}+\sin^2\theta_W)M^2_Z\cos2\beta\right)\delta_{ij}~, \nonumber\\
M^2_{\tilde{l}\,RR\,ij} &= m^2_{\tilde{E}\,ij} + \left(m^2_{l_i}-\sin^2\theta_WM^2_Z\cos2\beta\right)\delta_{ij}~, \nonumber\\
M^2_{\tilde{l}\,LR\,ij} &= v_1\mathcal{A}^l_{ij}-m_{l_i}\mu\tan\beta\delta_{ij}~. 
\end{align}
In these equations the indices $i,j \in \{1, 2, 3\}$ denote the three generations, $m_{l_i}$ are the lepton masses, 
$\theta_W$ is the weak mixing angle, $m_Z$ is the $Z$ boson mass, $\tan \beta=v_2/v_1$ with $v_1 = \left < H_1 \right >$ 
and $v_2 = \left < H_2 \right >$ being the two vacuum expectation values of the corresponding $SU(2)$ Higgs doublets, and 
$\mu$ is the Higgsino mass term.  Here, $\delta_{ij}$ is the Kronecker delta symbol.  The flavour violating terms in 
the $LL$ and $RR$ mixing matrices correspond to off-diagonal terms in the soft masses $m^2_{\tilde{L}\,ij}$ and 
$m^2_{\tilde{E}\,ij}$, respectively.

\begin{figure}[!tb]
\centering
\begin{subfigure}[b]{0.49\textwidth}
\centering
\includegraphics[width=\textwidth]{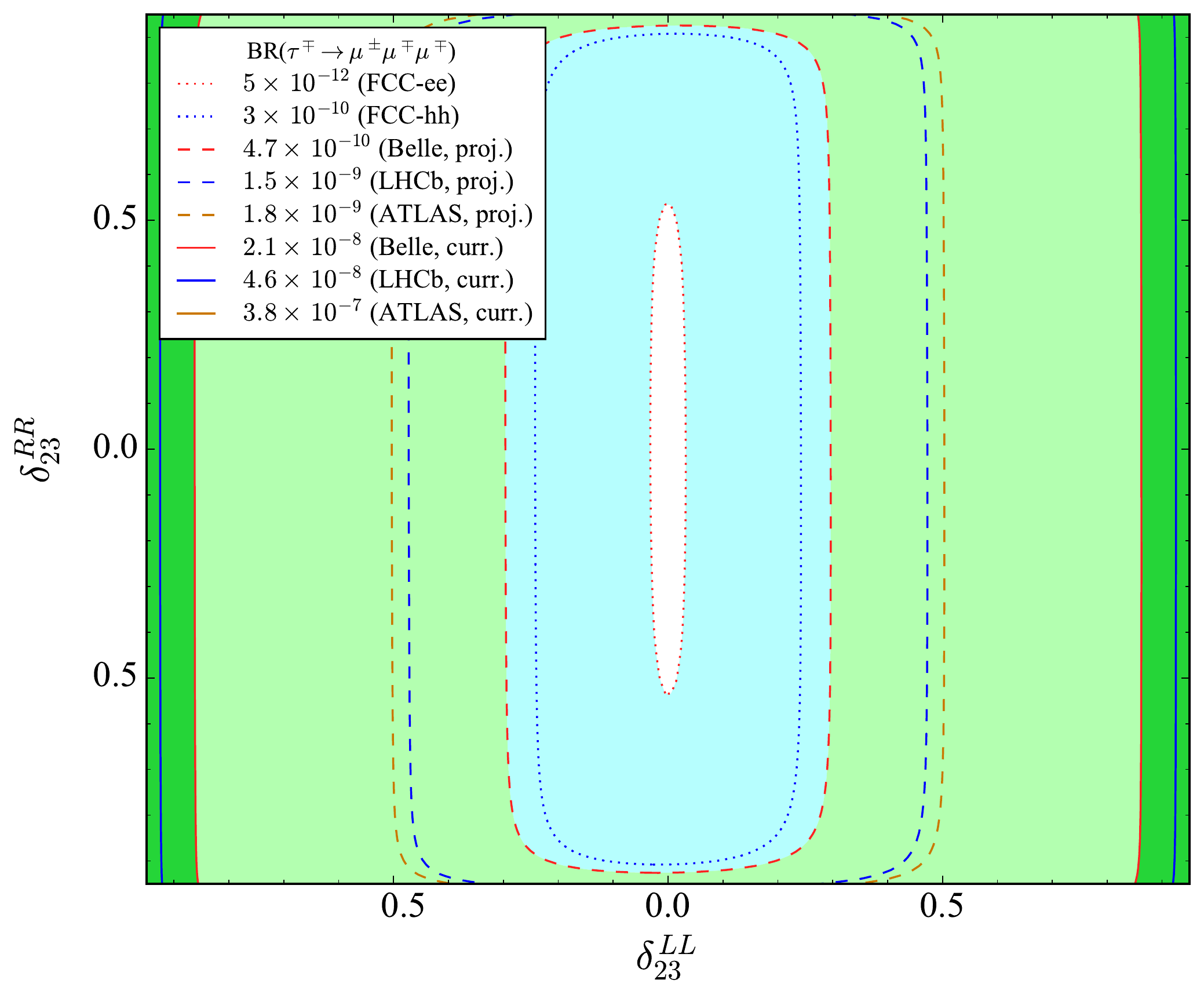}
\caption{}
\label{fig:tau_3muLL23RR23}
\end{subfigure}
\hfill
\begin{subfigure}[b]{0.49\textwidth}
\centering
\includegraphics[width=\textwidth]{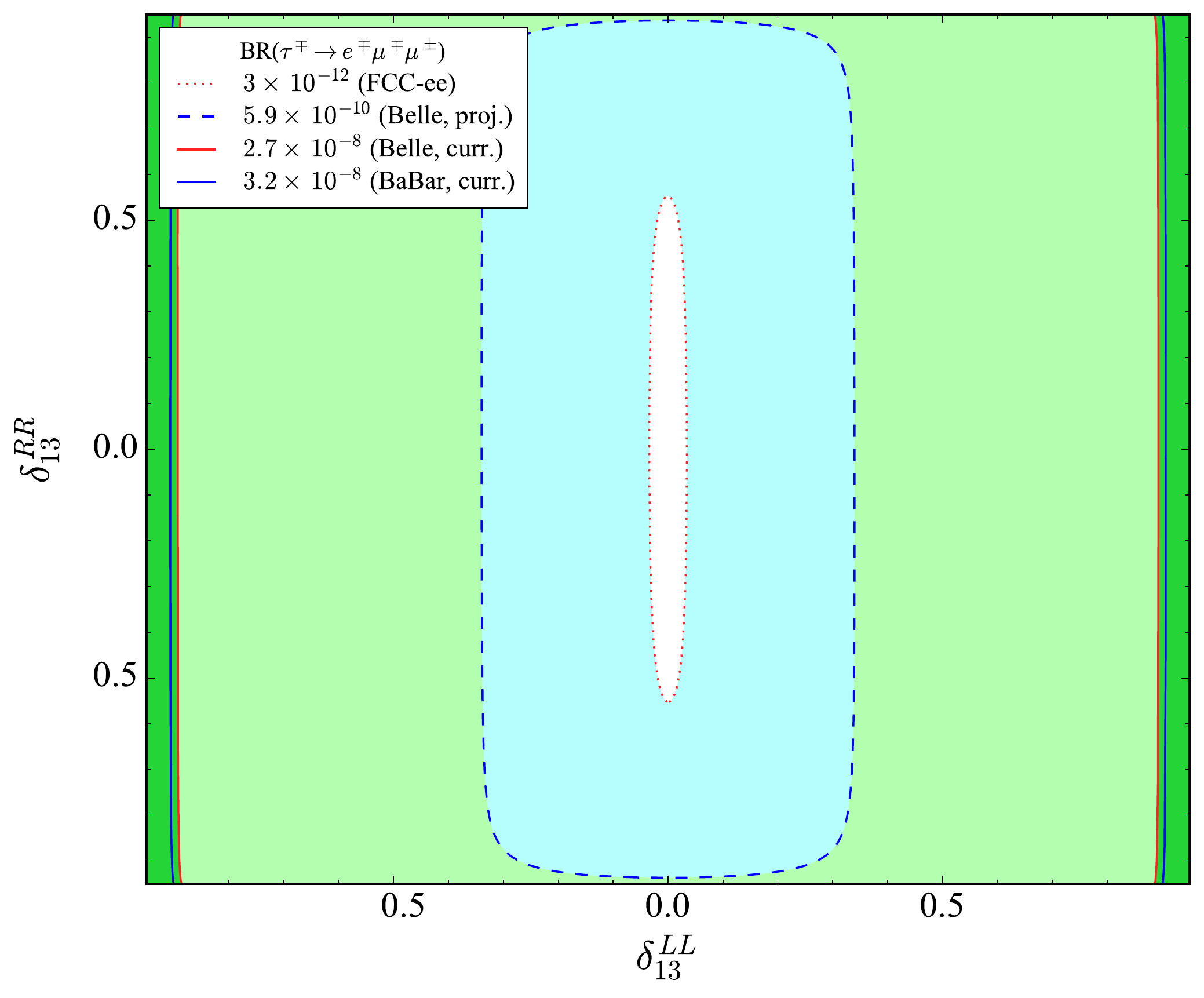}
\caption{}
\label{fig:tau_e2muLL13RR13}
\end{subfigure}
\hfill
\caption{Current and future branching ratio limits in the parameter plane of (a) $\delta^{LL}_{23}$ and $\delta^{RR}_{23}$ for the decay $\tau^{\mp} \to \mu^{\pm}\mu^{\mp}\mu^{\mp}$ and (b) $\delta^{LL}_{13}$ and $\delta^{RR}_{13}$ for the decay $\tau^\mp \to e^\mp \mu^\mp \mu^\pm$ in the MSSM.}
\label{fig:MSSM_one}
\end{figure}

Within the MSSM the sneutrino mass matrix has a one-block $3 \times 3$ form denoted as $\mathcal{M}^2_{\tilde{\nu}}$, where in the electroweak basis $(\tilde{\nu}_{eL},\tilde{\nu}_{\mu L},\tilde{\nu}_{\tau L})$, 
\begin{equation}
\mathcal{M}^2_{\tilde{\nu}} = M_{\tilde{\nu} \,LL}^2~, \qquad M^2_{\tilde{\nu}\,LL\,ij} = m^2_{\tilde{L}\,ij} + \left(\frac{1}{2}M^2_Z\cos2\beta\right)\delta_{ij}~.
\end{equation}
To parametrise the off-diagonal entries, we introduce the dimensionless real parameters,
\begin{equation}
\delta_{ij}^{AB} \equiv \frac{M^2_{\tilde{l}\,AB\,ij}}{m_{\tilde{A}_i} m_{\tilde{B}_j}}~,
\end{equation}
where $m_{\tilde{L}_i}$ and $m_{\tilde{E}_i}$ are the soft mass scales. We further assume that $|\delta_{ij}^{AB}| \leq 1$, and the hermiticity of $\mathcal{M}_{\tilde{l}}^2$ implies $\delta_{ij}^{AB} = \delta_{ji}^{BA}$.
After rotating the sleptons and sneutrinos into their mass eigenstates,
\begin{align} 
\mathrm{diag} \{m_{\tilde{l}_1}^2 ,m_{\tilde{l}_2}^2 ,m_{\tilde{l}_3}^2 ,m_{\tilde{l}_4}^2 ,m_{\tilde{l}_5}^2 ,m_{\tilde{l}_6}^2 \} &= R^{\tilde{l}} \mathcal{M}^2_{\tilde{l}} R^{\tilde{l}\dagger}~, \nonumber \\
\mathrm{diag} \{m_{\tilde{\nu}_1}^2 ,m_{\tilde{\nu}_2}^2 ,m_{\tilde{\nu}_3}^2  \} &= R^{\tilde{\nu}} \mathcal{M}^2_{\tilde{\nu}} R^{\tilde{\nu}\dagger}~, 
\end{align}
the soft breaking terms $m^2_{\tilde{L}\,ij}$, $m^2_{\tilde{E}\,ij}$ and $\mathcal{A}^l_{ij}$ can induce flavour-changing neutral current interactions, such as that between a lepton, slepton and neutralino, as shown in the Feynman diagram in Fig.~\ref{fig:mssm-tau_3mu-diagram}.

To numerically compute the impact of the present and future LFV constraints on the flavour violating parameters $\delta^{LL}_{ij}$ and $\delta^{RR}_{ij}$, we work with the following benchmark point for the MSSM parameters that provides a particle spectrum in agreement with the present collider limits:
\begin{align}
\tan\beta = 10~, &\qquad \mu = -100~\mathrm{GeV}~, \nonumber\\
M_A = 1000~\mathrm{GeV}~, &\qquad M_1 = 250~\mathrm{GeV}~, \nonumber\\
M_2 = 500~\mathrm{GeV}~, &\qquad M_3 = 2000~\mathrm{GeV}~, \nonumber\\
m_{\tilde{L}_i} = m_{\tilde{E}_j} = 1000~\mathrm{GeV}~, &\qquad A_{\tau} = 200~\mathrm{GeV}~.
\end{align}

We do not specify squark supersymmetry breaking parameters here, as their values are not relevant for the processes we calculate.
While searches for squarks and gluinos by ATLAS \cite{Aad:2015iea, Aad:2015pfx} and CMS \cite{Khachatryan:2016xdt,Khachatryan:2016epu} have pushed their respective mass limits to already rather large values, limits for slepton masses are still fairly weak \cite{Olive:2016xmw}. Direct slepton pair production requires the exchange of electroweak gauge bosons and is thus strongly suppressed compared to squark or gluino pair production at hadron colliders. Hence, assuming LFV is realised in nature, much stronger limits on the slepton masses can be obtained indirectly by measuring rare flavour violating lepton decays.

\begin{figure}[!tb]
\centering
\begin{subfigure}[b]{0.49\textwidth}
\centering
\includegraphics[width=\textwidth]{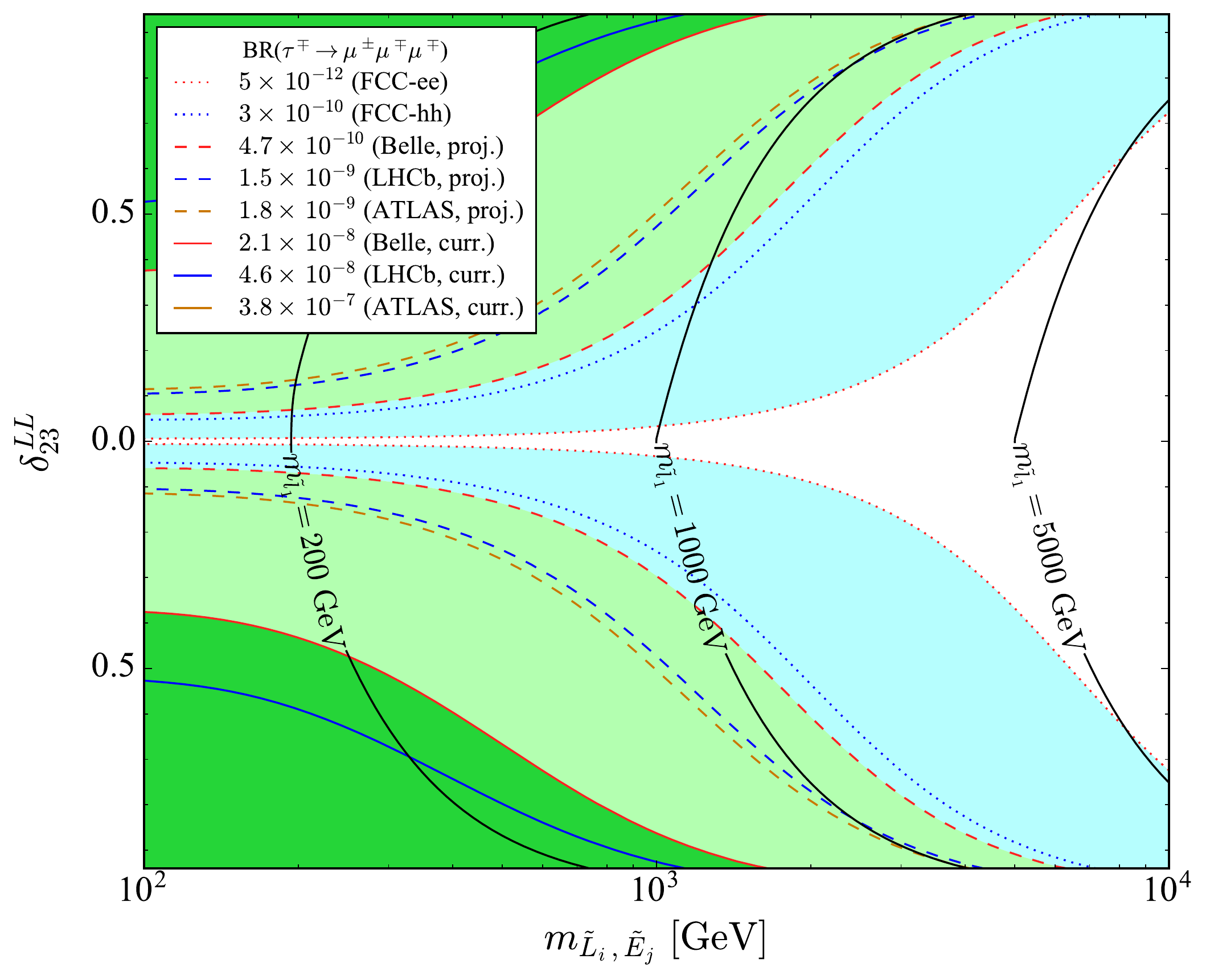}
\caption{}
\label{fig:tau_3musoftLL23}
\end{subfigure}
\hfill
\begin{subfigure}[b]{0.49\textwidth}
\centering
\includegraphics[width=\textwidth]{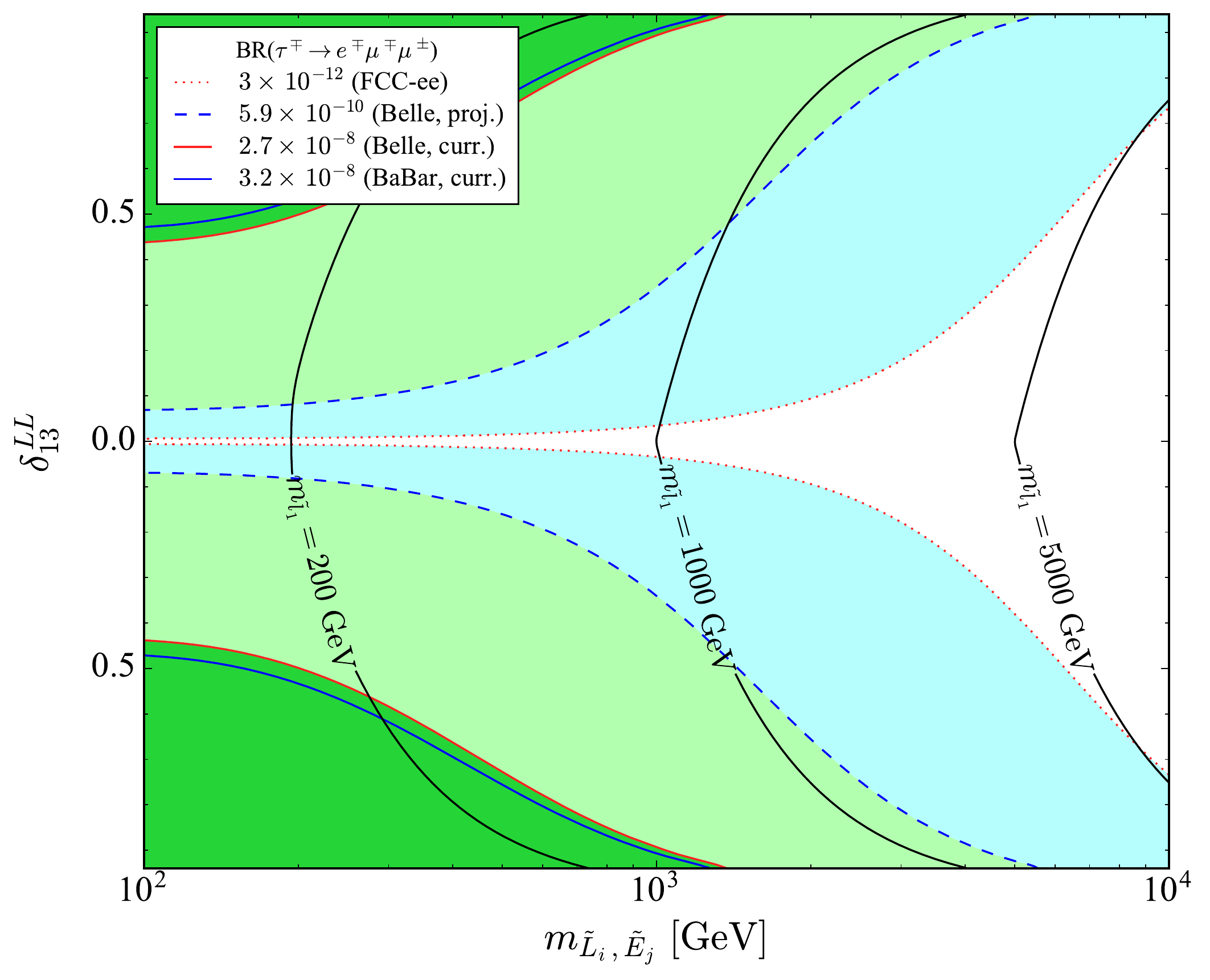}
\caption{}
\label{fig:tau_e2musoftLL13}
\end{subfigure}
\hfill
\caption{Current and future branching ratio limits in the parameter plane of (a) $m_{\tilde{L}_i\, ,\tilde{E}_j}$ and $\delta^{LL}_{23}$ for the decay $\tau^{\mp} \to \mu^{\pm}\mu^{\mp}\mu^{\mp}$ and (b) $m_{\tilde{L}_i\, ,\tilde{E}_j}$ and $\delta^{LL}_{13}$ for the decay $\tau^\mp \to e^\mp \mu^\mp \mu^\pm$ in the MSSM. The solid black lines represent constant values of the mass of the slepton $\tilde{l}_1$.}
\label{fig:MSSM_two}
\end{figure}

In Figs.~\ref{fig:tau_3muLL23RR23} and \ref{fig:tau_e2muLL13RR13}, we show present and future constraints on the pair $(\delta^{LL}_{23}, \delta^{RR}_{23})$ from the process $\tau^\mp \to \mu^\mp \mu^\mp \mu^\pm$, and the pair $(\delta^{LL}_{13}, \delta^{RR}_{13})$ from the process $\tau^\mp \to e^\mp \mu^\mp \mu^\pm$, respectively. In analogy with the squark sector \cite{Dittmaier:2007uw}, we find that the $\delta^{RR}_{13}$ and $\delta^{RR}_{23}$ parameters are much less constrained than their $LL$ counterparts. This is because the processes are mediated by flavour violating neutralino interactions. In the gauge-interaction basis, the exchanged particles are the bino ($\tilde{B}$), wino ($\tilde{W}^0$) or Higgsino ($\tilde{H}_i$) 
particles. The $\tilde{H}_i-l_R-\tilde{l}_L$ interactions are proportional to the lepton's Yukawa coupling $y_l$ and 
are thus subleading, while $\tilde{B}-l_{R/L}-\tilde{l}_{R/L}$ and $\tilde{W}^0-l_L-\tilde{l}_L$ interactions occur 
with the strength of their associated gauge couplings. Therefore, the branching ratios $\tau^\mp \to \mu^\mp \mu^\mp \mu^\pm$ and $\tau^\mp \to e^\mp \mu^\mp \mu^\pm$ are amplified for a light wino-type neutralino, i.e. small $M_2$, and large $\delta^{LL}_{ij}$.

In Fig.~\ref{fig:MSSM_two}, we show the LFV branching ratio limits where the soft slepton mass scale is allowed to 
vary along with a single mixing parameter. We vary the slepton mass scale over a wide range. For slepton masses at 
the current lower bound from direct searches ($ \sim 100$~GeV) future experiments could place very strong 
constraints on LFV parameters.  Since the 
slepton masses are large when the soft slepton mass scales $m_{\tilde{L}_i}=m_{\tilde{E}_j}$ are large, their contribution to LFV processes 
decouples and the sensitivity to the mixing parameters is reduced.

\section{Conclusions}\label{sec:conclusions}

The experimental observation of lepton flavour violation  would unambiguously serve as striking evidence for BSM 
physics, since in the SM lepton flavour violation is absent to all orders in perturbation theory. A plethora of the 
ongoing and near future experiments are likely to improve their sensitivity in the $\tau$ sector and will probe 
branching ratios at the level of $\mathcal{O}(10^{-10}-10^{-12})$.  

In this work we analyse the flavour violation in the $\tau$ sector, with a particular focus on the decays 
$\tau^{\mp} \to \mu^{\pm} \mu^{\mp} \mu^{\mp}$, $\tau^\mp \to e^\pm \mu^\mp \mu^\mp$ and 
$\tau^\mp \to e^\mp \mu^\mp \mu^\pm$ that can arise in various BSM models either at tree-level or with a loop suppression. We review the existing bounds on the branching ratio limits from Belle, BaBar and the LHC, and summarise the future sensitivity that these could achieve. We also discuss the limits that future circular colliders could reach. In the context of these limits, we provide an analysis of the parameter space that can be restricted in three BSM models that have lepton flavour violating interactions. Our findings are: 

\begin{itemize}

\item
The most stringent limit on the $\tau^{\mp} \to \mu^{\pm} \mu^{\mp} \mu^{\mp}$ decay is given by the Belle experiment, 
with an upper limit on the branching fraction equal to $2.1\times 10^{-8}$ at 90$\%$ C.L.  The LHCb experiment has 
produced an exclusion limit about two times larger.  In the near future the Belle-II experiment will extend
sensitivity down to a branching fraction of $4.7\times 10^{-10}$.  Although the present limit from ATLAS is an order 
of magnitude larger than the limit from Belle, the existing and upcoming 13 TeV data sets provide an opportunity for 
all of the LHC experiments to achieve better sensitivity than Belle.  These experiments could produce the
strongest limits for several years, until the Belle-II experiment analyses its full data set.  The future circular 
collider FCC-ee could further improve the limits down to $5 \times 10^{-12}$, an improvement of almost four orders 
of magnitude compared to the present bounds. For the $\tau^{\mp} \to e^{\pm} \mu^{\mp} \mu^{\mp}$ and $\tau^\mp \to e^\mp \mu^\mp \mu^\pm$ decays, a similar improvement on the present bounds can be achieved.

\item
For the Type-II Seesaw Model with a small triplet vev $v_{\Delta}$ in the range $(10^{-11}-10^{-9})$ GeV that 
naturally explains the $(0.01-1)$ eV light neutrino mass with $\mathcal{O}(1)$ Yukawa coupling $Y_{\Delta}$, the model parameter $\mu_{\Delta}$ is presently  constrained as $\mu_{\Delta} \geq (2\times 10^{-9}-7 \times 10^{-8})$ GeV. The future circular collider FCC-ee could provide improved constraints on $\mu_{\Delta}$ by almost two orders of magnitude. 
Constraints on the Dirac CP violating phase $\delta$ of the PMNS mixing matrix could be obtained by the Belle-II experiment in 
regions around $\pi /2$ and $3\pi/2$ for a quasi-degenerate neutrino spectrum with the oscillation angles equal to their best-fit values. 

\item
For the LRSM we consider two extreme regimes, with a lower and higher value of the symmetry breaking scale $v_R$ respectively. For the 
first benchmark point BP1, we consider a somewhat lower $v_R =8$ TeV and a large $\alpha_3 \sim \mathcal{O}(10)$, and for BP2, we consider a larger $v_R=30$ TeV with a smaller $\alpha_3 \sim \mathcal{O}(1)$, which is well within the perturbative regime. In BP1, and for a  doubly charged Higgs mass $M_{\delta^{\pm \pm}_R} = 800$ GeV, we find that the right-handed neutrino masses  $m_N \leq 290$ GeV are in agreement with the present stringent limit from  Belle. The future limits from LHCb and Belle-II  will further constrain the right-handed neutrino masses down to the $m_N \leq 100$ GeV mass range. Further improvements at the future circular colliders will allow for tighter constraints on the $\rho_2$ parameter and the doubly charged Higgs mass $M_{\delta^{\pm \pm}_R}$ to be obtained.

\item
Finally, for the MSSM, we explore the present and future constraints on the dimensionless LFV parameters $\delta^{LL}_{13}$, $\delta^{LL}_{23}$ (and their $RR$ equivalents) and the soft slepton masses from the $\tau^{\mp} \to \mu^{\mp} \mu^{\mp} \mu^{\pm}$ and $\tau^\mp \to e^\pm \mu^\mp \mu^\mp$ decays. We find that $\delta^{LL}_{13}$ and $\delta^{LL}_{23}$ are at present bounded by Belle to $|\delta^{LL}_{13,23}| \lesssim 0.9$ for the benchmark scenario we chose.  The future constraints 
from existing colliders will improve the limits to $\sim 0.2$, while an FCC-ee collider could further constrain this 
parameter to as low as $0.03$.  
\end{itemize} 

\vskip 1 \baselineskip

\acknowledgments
MS is supported in part by the European Commission through the ``HiggsTools'' Initial Training Network PITN-GA-2012-316704. MM is supported by the DST-INSPIRE grant INSPIRE-15-0074 and the Royal Society International Exchange Award.

  
\bibliographystyle{JHEP}


\bibliography{references}

\providecommand{\href}[2]{#2}\begingroup\raggedright\begin{thebibliography}{100}

\bibitem{Minkowski:1977sc}
P.~Minkowski, \emph{{$\mu \to e\gamma$ at a Rate of One Out of $10^{9}$ Muon
  Decays?}}, \href{http://dx.doi.org/10.1016/0370-2693(77)90435-X}{\emph{Phys.
  Lett.} {\bf B67} (1977) 421--428}.

\bibitem{Mohapatra:1979ia}
R.~N. Mohapatra and G.~Senjanovic, \emph{{Neutrino Mass and Spontaneous Parity
  Violation}}, \href{http://dx.doi.org/10.1103/PhysRevLett.44.912}{\emph{Phys.
  Rev. Lett.} {\bf 44} (1980) 912}.

\bibitem{Yanagida:1979as}
T.~Yanagida, \emph{{Horizontal Symmetry and Masses of Neutrinos}}, {\emph{Conf.
  Proc.} {\bf C7902131} (1979) 95--99}.

\bibitem{GellMann:1980vs}
M.~Gell-Mann, P.~Ramond and R.~Slansky, \emph{{Complex Spinors and Unified
  Theories}}, {\emph{Conf. Proc.} {\bf C790927} (1979) 315--321},
  [\href{https://arxiv.org/abs/1306.4669}{{\tt 1306.4669}}].

\bibitem{Schechter:1980gr}
J.~Schechter and J.~W.~F. Valle, \emph{{Neutrino Masses in SU(2) $\times$ U(1)
  Theories}}, \href{http://dx.doi.org/10.1103/PhysRevD.22.2227}{\emph{Phys.
  Rev.} {\bf D22} (1980) 2227}.

\bibitem{Magg:1980ut}
M.~Magg and C.~Wetterich, \emph{{Neutrino Mass Problem and Gauge Hierarchy}},
  \href{http://dx.doi.org/10.1016/0370-2693(80)90825-4}{\emph{Phys. Lett.} {\bf
  B94} (1980) 61--64}.

\bibitem{Lazarides:1980nt}
G.~Lazarides, Q.~Shafi and C.~Wetterich, \emph{{Proton Lifetime and Fermion
  Masses in an SO(10) Model}},
  \href{http://dx.doi.org/10.1016/0550-3213(81)90354-0}{\emph{Nucl. Phys.} {\bf
  B181} (1981) 287--300}.

\bibitem{Cheng:1980qt}
T.~P. Cheng and L.-F. Li, \emph{{Neutrino Masses, Mixings and Oscillations in
  SU(2) $\times$ U(1) Models of Electroweak Interactions}},
  \href{http://dx.doi.org/10.1103/PhysRevD.22.2860}{\emph{Phys. Rev.} {\bf D22}
  (1980) 2860}.

\bibitem{Mohapatra:1980yp}
R.~N. Mohapatra and G.~Senjanovic, \emph{{Neutrino Masses and Mixings in Gauge
  Models with Spontaneous Parity Violation}},
  \href{http://dx.doi.org/10.1103/PhysRevD.23.165}{\emph{Phys. Rev.} {\bf D23}
  (1981) 165}.

\bibitem{Foot:1988aq}
R.~Foot, H.~Lew, X.~G. He and G.~C. Joshi, \emph{{Seesaw Neutrino Masses
  Induced by a Triplet of Leptons}},
  \href{http://dx.doi.org/10.1007/BF01415558}{\emph{Z. Phys.} {\bf C44} (1989)
  441}.

\bibitem{Mohapatra:1986aw}
R.~N. Mohapatra, \emph{{Mechanism for Understanding Small Neutrino Mass in
  Superstring Theories}},
  \href{http://dx.doi.org/10.1103/PhysRevLett.56.561}{\emph{Phys. Rev. Lett.}
  {\bf 56} (1986) 561--563}.

\bibitem{Mohapatra:1986bd}
R.~N. Mohapatra and J.~W.~F. Valle, \emph{{Neutrino Mass and Baryon Number
  Nonconservation in Superstring Models}},
  \href{http://dx.doi.org/10.1103/PhysRevD.34.1642}{\emph{Phys. Rev.} {\bf D34}
  (1986) 1642}.

\bibitem{Wyler:1982dd}
D.~Wyler and L.~Wolfenstein, \emph{{Massless Neutrinos in Left-Right Symmetric
  Models}}, \href{http://dx.doi.org/10.1016/0550-3213(83)90482-0}{\emph{Nucl.
  Phys.} {\bf B218} (1983) 205--214}.

\bibitem{Witten:1985bz}
E.~Witten, \emph{{New Issues in Manifolds of SU(3) Holonomy}},
  \href{http://dx.doi.org/10.1016/0550-3213(86)90202-6}{\emph{Nucl. Phys.} {\bf
  B268} (1986) 79}.

\bibitem{Hewett:1988xc}
J.~L. Hewett and T.~G. Rizzo, \emph{{Low-Energy Phenomenology of Superstring
  Inspired E(6) Models}},
  \href{http://dx.doi.org/10.1016/0370-1573(89)90071-9}{\emph{Phys. Rept.} {\bf
  183} (1989) 193}.

\bibitem{Pati:1974yy}
J.~C. Pati and A.~Salam, \emph{{Lepton Number as the Fourth Color}},
  \href{http://dx.doi.org/10.1103/PhysRevD.10.275,
  10.1103/PhysRevD.11.703.2}{\emph{Phys. Rev.} {\bf D10} (1974) 275--289}.

\bibitem{Mohapatra:1974gc}
R.~N. Mohapatra and J.~C. Pati, \emph{{A Natural Left-Right Symmetry}},
  \href{http://dx.doi.org/10.1103/PhysRevD.11.2558}{\emph{Phys. Rev.} {\bf D11}
  (1975) 2558}.

\bibitem{Senjanovic:1975rk}
G.~Senjanovic and R.~N. Mohapatra, \emph{{Exact Left-Right Symmetry and
  Spontaneous Violation of Parity}},
  \href{http://dx.doi.org/10.1103/PhysRevD.12.1502}{\emph{Phys. Rev.} {\bf D12}
  (1975) 1502}.

\bibitem{Duka:1999uc}
P.~Duka, J.~Gluza and M.~Zralek, \emph{{Quantization and renormalization of the
  manifest left-right symmetric model of electroweak interactions}},
  \href{http://dx.doi.org/10.1006/aphy.1999.5988}{\emph{Annals Phys.} {\bf 280}
  (2000) 336--408}, [\href{https://arxiv.org/abs/hep-ph/9910279}{{\tt
  hep-ph/9910279}}].

\bibitem{Nilles:1983ge}
H.~P. Nilles, \emph{{Supersymmetry, Supergravity and Particle Physics}},
  \href{http://dx.doi.org/10.1016/0370-1573(84)90008-5}{\emph{Phys. Rept.} {\bf
  110} (1984) 1--162}.

\bibitem{Haber:1984rc}
H.~E. Haber and G.~L. Kane, \emph{{The Search for Supersymmetry: Probing
  Physics Beyond the Standard Model}},
  \href{http://dx.doi.org/10.1016/0370-1573(85)90051-1}{\emph{Phys. Rept.} {\bf
  117} (1985) 75--263}.

\bibitem{Martin:1997ns}
S.~P. Martin, \emph{{A Supersymmetry primer}},
  \href{https://arxiv.org/abs/hep-ph/9709356}{{\tt hep-ph/9709356}}.

\bibitem{Weinberg:1979sa}
S.~Weinberg, \emph{{Baryon and Lepton Nonconserving Processes}},
  \href{http://dx.doi.org/10.1103/PhysRevLett.43.1566}{\emph{Phys. Rev. Lett.}
  {\bf 43} (1979) 1566--1570}.

\bibitem{Wilczek:1979hc}
F.~Wilczek and A.~Zee, \emph{{Operator Analysis of Nucleon Decay}},
  \href{http://dx.doi.org/10.1103/PhysRevLett.43.1571}{\emph{Phys. Rev. Lett.}
  {\bf 43} (1979) 1571--1573}.

\bibitem{Adam:2013mnn}
{\scshape MEG} collaboration, J.~Adam et~al., \emph{{New constraint on the
  existence of the $\mu^+ \to e^+\gamma$ decay}},
  \href{http://dx.doi.org/10.1103/PhysRevLett.110.201801}{\emph{Phys. Rev.
  Lett.} {\bf 110} (2013) 201801}, [\href{https://arxiv.org/abs/1303.0754}{{\tt
  1303.0754}}].

\bibitem{Olive:2016xmw}
{\scshape Particle Data Group} collaboration, C.~Patrignani et~al.,
  \emph{{Review of Particle Physics}},
  \href{http://dx.doi.org/10.1088/1674-1137/40/10/100001}{\emph{Chin. Phys.}
  {\bf C40} (2016) 100001}.

\bibitem{Bellgardt:1987du}
{\scshape SINDRUM} collaboration, U.~Bellgardt et~al., \emph{{Search for the
  Decay mu+ $\to$ e+ e+ e-}},
  \href{http://dx.doi.org/10.1016/0550-3213(88)90462-2}{\emph{Nucl. Phys.} {\bf
  B299} (1988) 1--6}.

\bibitem{Khachatryan:2015kon}
{\scshape CMS} collaboration, V.~Khachatryan et~al., \emph{{Search for
  Lepton-Flavour-Violating Decays of the Higgs Boson}},
  \href{http://dx.doi.org/10.1016/j.physletb.2015.07.053}{\emph{Phys. Lett.}
  {\bf B749} (2015) 337--362}, [\href{https://arxiv.org/abs/1502.07400}{{\tt
  1502.07400}}].

\bibitem{Aad:2015gha}
{\scshape ATLAS} collaboration, G.~Aad et~al., \emph{{Search for
  lepton-flavour-violating $H\to\mu\tau$ decays of the Higgs boson with the
  ATLAS detector}},
  \href{http://dx.doi.org/10.1007/JHEP11(2015)211}{\emph{JHEP} {\bf 11} (2015)
  211}, [\href{https://arxiv.org/abs/1508.03372}{{\tt 1508.03372}}].

\bibitem{Dassinger:2007ru}
B.~M. Dassinger, T.~Feldmann, T.~Mannel and S.~Turczyk,
  \emph{{Model-independent analysis of lepton flavour violating tau decays}},
  \href{http://dx.doi.org/10.1088/1126-6708/2007/10/039}{\emph{JHEP} {\bf 10}
  (2007) 039}, [\href{https://arxiv.org/abs/0707.0988}{{\tt 0707.0988}}].

\bibitem{Harnik:2012pb}
R.~Harnik, J.~Kopp and J.~Zupan, \emph{{Flavor Violating Higgs Decays}},
  \href{http://dx.doi.org/10.1007/JHEP03(2013)026}{\emph{JHEP} {\bf 03} (2013)
  026}, [\href{https://arxiv.org/abs/1209.1397}{{\tt 1209.1397}}].

\bibitem{Falkowski:2013jya}
A.~Falkowski, D.~M. Straub and A.~Vicente, \emph{{Vector-like leptons: Higgs
  decays and collider phenomenology}},
  \href{http://dx.doi.org/10.1007/JHEP05(2014)092}{\emph{JHEP} {\bf 05} (2014)
  092}, [\href{https://arxiv.org/abs/1312.5329}{{\tt 1312.5329}}].

\bibitem{Heeck:2014qea}
J.~Heeck, M.~Holthausen, W.~Rodejohann and Y.~Shimizu, \emph{{Higgs $\to
  \mu\tau$ in Abelian and non-Abelian flavor symmetry models}},
  \href{http://dx.doi.org/10.1016/j.nuclphysb.2015.04.025}{\emph{Nucl. Phys.}
  {\bf B896} (2015) 281--310}, [\href{https://arxiv.org/abs/1412.3671}{{\tt
  1412.3671}}].

\bibitem{Crivellin:2015mga}
A.~Crivellin, G.~D'Ambrosio and J.~Heeck, \emph{{Explaining
  $h\to\mu^\pm\tau^\mp$, $B\to K^* \mu^+\mu^-$ and $B\to K \mu^+\mu^-/B\to K
  e^+e^-$ in a two-Higgs-doublet model with gauged $L_\mu-L_\tau$}},
  \href{http://dx.doi.org/10.1103/PhysRevLett.114.151801}{\emph{Phys. Rev.
  Lett.} {\bf 114} (2015) 151801},
  [\href{https://arxiv.org/abs/1501.00993}{{\tt 1501.00993}}].

\bibitem{Banerjee:2016foh}
S.~Banerjee, B.~Bhattacherjee, M.~Mitra and M.~Spannowsky, \emph{{The Lepton
  Flavour Violating Higgs Decays at the HL-LHC and the ILC}},
  \href{http://dx.doi.org/10.1007/JHEP07(2016)059}{\emph{JHEP} {\bf 07} (2016)
  059}, [\href{https://arxiv.org/abs/1603.05952}{{\tt 1603.05952}}].

\bibitem{Chakraborty:2016gff}
I.~Chakraborty, A.~Datta and A.~Kundu, \emph{{Lepton flavor violating Higgs
  boson decay ${\boldsymbol{h}} \rightarrow \mu \tau $ at the ILC}},
  \href{http://dx.doi.org/10.1088/0954-3899/43/12/125001}{\emph{J. Phys.} {\bf
  G43} (2016) 125001}, [\href{https://arxiv.org/abs/1603.06681}{{\tt
  1603.06681}}].

\bibitem{Hirsch:1996qw}
M.~Hirsch, H.~V. Klapdor-Kleingrothaus and O.~Panella, \emph{{Double beta decay
  in left-right symmetric models}},
  \href{http://dx.doi.org/10.1016/0370-2693(96)00185-2}{\emph{Phys. Lett.} {\bf
  B374} (1996) 7--12}, [\href{https://arxiv.org/abs/hep-ph/9602306}{{\tt
  hep-ph/9602306}}].

\bibitem{Barry:2013xxa}
J.~Barry and W.~Rodejohann, \emph{{Lepton number and flavour violation in
  TeV-scale left-right symmetric theories with large left-right mixing}},
  \href{http://dx.doi.org/10.1007/JHEP09(2013)153}{\emph{JHEP} {\bf 09} (2013)
  153}, [\href{https://arxiv.org/abs/1303.6324}{{\tt 1303.6324}}].

\bibitem{Awasthi:2013ff}
R.~L. Awasthi, M.~K. Parida and S.~Patra, \emph{{Neutrino masses, dominant
  neutrinoless double beta decay, and observable lepton flavor violation in
  left-right models and SO(10) grand unification with low mass $ W_R, Z_R$
  bosons}}, \href{http://dx.doi.org/10.1007/JHEP08(2013)122}{\emph{JHEP} {\bf
  08} (2013) 122}, [\href{https://arxiv.org/abs/1302.0672}{{\tt 1302.0672}}].

\bibitem{Bambhaniya:2015ipg}
G.~Bambhaniya, P.~S.~B. Dev, S.~Goswami and M.~Mitra, \emph{{The Scalar Triplet
  Contribution to Lepton Flavour Violation and Neutrinoless Double Beta Decay
  in Left-Right Symmetric Model}},
  \href{http://dx.doi.org/10.1007/JHEP04(2016)046}{\emph{JHEP} {\bf 04} (2016)
  046}, [\href{https://arxiv.org/abs/1512.00440}{{\tt 1512.00440}}].

\bibitem{Bonilla:2016fqd}
C.~Bonilla, M.~E. Krauss, T.~Opferkuch and W.~Porod, \emph{{Perspectives for
  Detecting Lepton Flavour Violation in Left-Right Symmetric Models}},
  \href{https://arxiv.org/abs/1611.07025}{{\tt 1611.07025}}.

\bibitem{Arganda:2008jj}
E.~Arganda, M.~J. Herrero and J.~Portoles, \emph{{Lepton flavour violating
  semileptonic tau decays in constrained MSSM-seesaw scenarios}},
  \href{http://dx.doi.org/10.1088/1126-6708/2008/06/079}{\emph{JHEP} {\bf 06}
  (2008) 079}, [\href{https://arxiv.org/abs/0803.2039}{{\tt 0803.2039}}].

\bibitem{Arana-Catania:2013ggc}
M.~Arana-Catania, S.~Heinemeyer and M.~J. Herrero, \emph{{New Constraints on
  General Slepton Flavor Mixing}},
  \href{http://dx.doi.org/10.1103/PhysRevD.88.015026}{\emph{Phys. Rev.} {\bf
  D88} (2013) 015026}, [\href{https://arxiv.org/abs/1304.2783}{{\tt
  1304.2783}}].

\bibitem{Arana-Catania:2013xma}
M.~Arana-Catania, E.~Arganda and M.~J. Herrero, \emph{{Non-decoupling SUSY in
  LFV Higgs decays: a window to new physics at the LHC}},
  \href{http://dx.doi.org/10.1007/JHEP10(2015)192,
  10.1007/JHEP09(2013)160}{\emph{JHEP} {\bf 09} (2013) 160},
  [\href{https://arxiv.org/abs/1304.3371}{{\tt 1304.3371}}].

\bibitem{Hayasaka:2010np}
K.~Hayasaka et~al., \emph{{Search for Lepton Flavor Violating Tau Decays into
  Three Leptons with 719 Million Produced Tau+Tau- Pairs}},
  \href{http://dx.doi.org/10.1016/j.physletb.2010.03.037}{\emph{Phys. Lett.}
  {\bf B687} (2010) 139--143}, [\href{https://arxiv.org/abs/1001.3221}{{\tt
  1001.3221}}].

\bibitem{babar}
{\scshape BaBar} collaboration, J.~P. Lees et~al., \emph{{Limits on tau
  Lepton-Flavor Violating Decays in three charged leptons}},
  \href{http://dx.doi.org/10.1103/PhysRevD.81.111101}{\emph{Phys. Rev.} {\bf
  D81} (2010) 111101}, [\href{https://arxiv.org/abs/1002.4550}{{\tt
  1002.4550}}].

\bibitem{Aaij:2014azz}
{\scshape LHCb} collaboration, R.~Aaij et~al., \emph{{Search for the lepton
  flavour violating decay $\tau^{-} \to \mu^{-} \mu^{+} \mu^{-}$}},
  \href{http://dx.doi.org/10.1007/JHEP02(2015)121}{\emph{JHEP} {\bf 02} (2015)
  121}, [\href{https://arxiv.org/abs/1409.8548}{{\tt 1409.8548}}].

\bibitem{Aad:2016wce}
{\scshape ATLAS} collaboration, G.~Aad et~al., \emph{{Probing lepton flavour
  violation via neutrinoless $\tau \longrightarrow 3\mu $ decays with the ATLAS
  detector}},
  \href{http://dx.doi.org/10.1140/epjc/s10052-016-4041-9}{\emph{Eur. Phys. J.}
  {\bf C76} (2016) 232}, [\href{https://arxiv.org/abs/1601.03567}{{\tt
  1601.03567}}].

\bibitem{belleprojection}
T.~Aushev et~al., \emph{{Physics at Super B Factory}},
  \href{https://arxiv.org/abs/1002.5012}{{\tt 1002.5012}}.

\bibitem{Bediaga:2012py}
{\scshape LHCb} collaboration, R.~Aaij et~al., \emph{{Implications of LHCb
  measurements and future prospects}},
  \href{http://dx.doi.org/10.1140/epjc/s10052-013-2373-2}{\emph{Eur. Phys. J.}
  {\bf C73} (2013) 2373}, [\href{https://arxiv.org/abs/1208.3355}{{\tt
  1208.3355}}].

\bibitem{Aaij:2011jh}
{\scshape LHCb} collaboration, R.~Aaij et~al., \emph{{Measurement of $J/\psi$
  production in $pp$ collisions at $\sqrt{s}=7~\rm{TeV}$}},
  \href{http://dx.doi.org/10.1140/epjc/s10052-011-1645-y}{\emph{Eur. Phys. J.}
  {\bf C71} (2011) 1645}, [\href{https://arxiv.org/abs/1103.0423}{{\tt
  1103.0423}}].

\bibitem{Aaij:2015rla}
{\scshape LHCb} collaboration, R.~Aaij et~al., \emph{{Measurement of forward
  $J/\psi$ production cross-sections in $pp$ collisions at $\sqrt{s}=13$ TeV}},
  \href{http://dx.doi.org/10.1007/JHEP10(2015)172}{\emph{JHEP} {\bf 10} (2015)
  172}, [\href{https://arxiv.org/abs/1509.00771}{{\tt 1509.00771}}].

\bibitem{Aaij:2013mga}
{\scshape LHCb} collaboration, R.~Aaij et~al., \emph{{Prompt charm production
  in pp collisions at sqrt(s)=7 TeV}},
  \href{http://dx.doi.org/10.1016/j.nuclphysb.2013.02.010}{\emph{Nucl. Phys.}
  {\bf B871} (2013) 1--20}, [\href{https://arxiv.org/abs/1302.2864}{{\tt
  1302.2864}}].

\bibitem{Aaij:2015bpa}
{\scshape LHCb} collaboration, R.~Aaij et~al., \emph{{Measurements of prompt
  charm production cross-sections in $pp$ collisions at $ \sqrt{s}=13 $ TeV}},
  \href{http://dx.doi.org/10.1007/JHEP03(2016)159, 10.1007/JHEP09(2016)013,
  10.1007/JHEP03(2016)159 10.1007/JHEP09(2016)013}{\emph{JHEP} {\bf 03} (2016)
  159}, [\href{https://arxiv.org/abs/1510.01707}{{\tt 1510.01707}}].

\bibitem{Chatrchyan:2014mua}
{\scshape CMS} collaboration, S.~Chatrchyan et~al., \emph{{Measurement of
  inclusive W and Z boson production cross sections in pp collisions at
  $\sqrt{s}$ = 8 TeV}},
  \href{http://dx.doi.org/10.1103/PhysRevLett.112.191802}{\emph{Phys. Rev.
  Lett.} {\bf 112} (2014) 191802}, [\href{https://arxiv.org/abs/1402.0923}{{\tt
  1402.0923}}].

\bibitem{Aad:2016naf}
{\scshape ATLAS} collaboration, G.~Aad et~al., \emph{{Measurement of $W^{\pm}$
  and $Z$-boson production cross sections in $pp$ collisions at $\sqrt{s}=13$
  TeV with the ATLAS detector}},
  \href{http://dx.doi.org/10.1016/j.physletb.2016.06.023}{\emph{Phys. Lett.}
  {\bf B759} (2016) 601--621}, [\href{https://arxiv.org/abs/1603.09222}{{\tt
  1603.09222}}].

\bibitem{FCC}
M.~Benedikt and F.~Zimmermann, \emph{{Future Circular Colliders}},
  {\emph{CERN-ACC-2015-164} (2015) }.

\bibitem{Mangano:2016jyj}
M.~L. Mangano et~al., \emph{{Physics at a 100 TeV pp collider: Standard Model
  processes}},  \href{https://arxiv.org/abs/1607.01831}{{\tt 1607.01831}}.

\bibitem{dEnterria:2016sca}
D.~d'Enterria, \emph{{Physics at the FCC-ee}},  in \emph{{17th Lomonosov
  Conference on Elementary Particle Physics Moscow, Russia, August 20-26,
  2015}}, 2016.
\newblock \href{https://arxiv.org/abs/1602.05043}{{\tt 1602.05043}}.

\bibitem{Alwall2014}
J.~Alwall et~al., \emph{{The automated computation of tree-level and
  next-to-leading order differential cross sections, and their matching to
  parton shower simulations}},
  \href{http://dx.doi.org/10.1007/JHEP07(2014)079}{\emph{Journal of High Energy
  Physics} {\bf 2014} (2014) 79}, [\href{https://arxiv.org/abs/1405.0301}{{\tt
  1405.0301}}].

\bibitem{Alloul20142250}
A.~Alloul, N.~D. Christensen, C.~Degrande, C.~Duhr and B.~Fuks,
  \emph{{FeynRules  2.0 — A complete toolbox for tree-level
  phenomenology}},
  \href{http://dx.doi.org/http://dx.doi.org/10.1016/j.cpc.2014.04.012}{\emph{Computer
  Physics Communications} {\bf 185} (2014) 2250 -- 2300},
  [\href{https://arxiv.org/abs/1310.1921}{{\tt 1310.1921}}].

\bibitem{POROD2003275}
W.~Porod, \emph{{SPheno, a program for calculating supersymmetric spectra, SUSY
  particle decays and SUSY particle production at e+e− colliders}},
  \href{http://dx.doi.org/http://dx.doi.org/10.1016/S0010-4655(03)00222-4}{\emph{Computer
  Physics Communications} {\bf 153} (2003) 275 -- 315},
  [\href{https://arxiv.org/abs/hep-ph/0301101}{{\tt hep-ph/0301101}}].

\bibitem{Porod20122458}
W.~Porod and F.~Staub, \emph{{SPheno 3.1: extensions including flavour,
  CP-phases and models beyond the MSSM}},
  \href{http://dx.doi.org/http://dx.doi.org/10.1016/j.cpc.2012.05.021}{\emph{Computer
  Physics Communications} {\bf 183} (2012) 2458 -- 2469},
  [\href{https://arxiv.org/abs/1104.1573}{{\tt 1104.1573}}].

\bibitem{Staub20141773}
F.~Staub, \emph{{SARAH   4: A tool for (not only SUSY) model builders}},
  \href{http://dx.doi.org/http://dx.doi.org/10.1016/j.cpc.2014.02.018}{\emph{Computer
  Physics Communications} {\bf 185} (2014) 1773 -- 1790},
  [\href{https://arxiv.org/abs/1309.7223}{{\tt 1309.7223}}].

\bibitem{ATLAS-CONF-2016-051}
{\scshape ATLAS} collaboration, G.~Aad et~al., \emph{{Search for doubly-charged
  Higgs bosons in same-charge electron pair final states using proton-proton
  collisions at $\sqrt{s}=13\,\mathrm{TeV}$ with the ATLAS detector}},  Tech.
  Rep. ATLAS-CONF-2016-051, CERN, Geneva, Aug, 2016.

\bibitem{Aad:2015xaa}
{\scshape ATLAS} collaboration, G.~Aad et~al., \emph{{Search for heavy Majorana
  neutrinos with the ATLAS detector in pp collisions at $ \sqrt{s}=8 $ TeV}},
  \href{http://dx.doi.org/10.1007/JHEP07(2015)162}{\emph{JHEP} {\bf 07} (2015)
  162}, [\href{https://arxiv.org/abs/1506.06020}{{\tt 1506.06020}}].

\bibitem{Khachatryan:2014dka}
{\scshape CMS} collaboration, V.~Khachatryan et~al., \emph{{Search for heavy
  neutrinos and $\mathrm {W}$ bosons with right-handed couplings in
  proton-proton collisions at $\sqrt{s} = 8\,\text {TeV} $}},
  \href{http://dx.doi.org/10.1140/epjc/s10052-014-3149-z}{\emph{Eur. Phys. J.}
  {\bf C74} (2014) 3149}, [\href{https://arxiv.org/abs/1407.3683}{{\tt
  1407.3683}}].

\bibitem{ATLAS:2015nsi}
{\scshape ATLAS} collaboration, G.~Aad et~al., \emph{{Search for new phenomena
  in dijet mass and angular distributions from $pp$ collisions at $\sqrt{s}=$
  13 TeV with the ATLAS detector}},
  \href{http://dx.doi.org/10.1016/j.physletb.2016.01.032}{\emph{Phys. Lett.}
  {\bf B754} (2016) 302--322}, [\href{https://arxiv.org/abs/1512.01530}{{\tt
  1512.01530}}].

\bibitem{Khachatryan:2015dcf}
{\scshape CMS} collaboration, V.~Khachatryan et~al., \emph{{Search for narrow
  resonances decaying to dijets in proton-proton collisions at $\sqrt(s) =$ 13
  TeV}}, \href{http://dx.doi.org/10.1103/PhysRevLett.116.071801}{\emph{Phys.
  Rev. Lett.} {\bf 116} (2016) 071801},
  [\href{https://arxiv.org/abs/1512.01224}{{\tt 1512.01224}}].

\bibitem{Deppisch:2015qwa}
F.~F. Deppisch, P.~S. Bhupal~Dev and A.~Pilaftsis, \emph{{Neutrinos and
  Collider Physics}},
  \href{http://dx.doi.org/10.1088/1367-2630/17/7/075019}{\emph{New J. Phys.}
  {\bf 17} (2015) 075019}, [\href{https://arxiv.org/abs/1502.06541}{{\tt
  1502.06541}}].

\bibitem{Babu:2016rcr}
K.~S. Babu and S.~Jana, \emph{{Probing Doubly Charged Higgs Bosons at the LHC
  through Photon Initiated Processes}},
  \href{https://arxiv.org/abs/1612.09224}{{\tt 1612.09224}}.

\bibitem{Nemevsek:2011hz}
M.~Nemevsek, F.~Nesti, G.~Senjanovic and Y.~Zhang, \emph{{First Limits on
  Left-Right Symmetry Scale from LHC Data}},
  \href{http://dx.doi.org/10.1103/PhysRevD.83.115014}{\emph{Phys. Rev.} {\bf
  D83} (2011) 115014}, [\href{https://arxiv.org/abs/1103.1627}{{\tt
  1103.1627}}].

\bibitem{Nemevsek:2016enw}
M.~Nemevšek, F.~Nesti and J.~C. Vasquez, \emph{{Majorana Higgses at
  colliders}},  \href{https://arxiv.org/abs/1612.06840}{{\tt 1612.06840}}.

\bibitem{Mitra:2016wpr}
M.~Mitra, S.~Niyogi and M.~Spannowsky, \emph{{Type-II Seesaw and Multilepton
  Signatures at Hadron Colliders}},
  \href{https://arxiv.org/abs/1611.09594}{{\tt 1611.09594}}.

\bibitem{Mitra:2016kov}
M.~Mitra, R.~Ruiz, D.~J. Scott and M.~Spannowsky, \emph{{Neutrino Jets from
  High-Mass $W_R$ Gauge Bosons in TeV-Scale Left-Right Symmetric Models}},
  \href{http://dx.doi.org/10.1103/PhysRevD.94.095016}{\emph{Phys. Rev.} {\bf
  D94} (2016) 095016}, [\href{https://arxiv.org/abs/1607.03504}{{\tt
  1607.03504}}].

\bibitem{Mattelaer:2016ynf}
O.~Mattelaer, M.~Mitra and R.~Ruiz, \emph{{Automated Neutrino Jet and Top Jet
  Predictions at Next-to-Leading-Order with Parton Shower Matching in Effective
  Left-Right Symmetric Models}},  \href{https://arxiv.org/abs/1610.08985}{{\tt
  1610.08985}}.

\bibitem{Lindner:2016lxq}
M.~Lindner, F.~S. Queiroz, W.~Rodejohann and C.~E. Yaguna, \emph{{Left-Right
  Symmetry and Lepton Number Violation at the Large Hadron Electron Collider}},
  \href{http://dx.doi.org/10.1007/JHEP06(2016)140}{\emph{JHEP} {\bf 06} (2016)
  140}, [\href{https://arxiv.org/abs/1604.08596}{{\tt 1604.08596}}].

\bibitem{Lindner:2016lpp}
M.~Lindner, F.~S. Queiroz and W.~Rodejohann, \emph{{Dilepton bounds on
  left–right symmetry at the LHC run II and neutrinoless double beta decay}},
  \href{http://dx.doi.org/10.1016/j.physletb.2016.08.068}{\emph{Phys. Lett.}
  {\bf B762} (2016) 190--195}, [\href{https://arxiv.org/abs/1604.07419}{{\tt
  1604.07419}}].

\bibitem{Melfo:2011nx}
A.~Melfo, M.~Nemevsek, F.~Nesti, G.~Senjanovic and Y.~Zhang, \emph{{Type II
  Seesaw at LHC: The Roadmap}},
  \href{http://dx.doi.org/10.1103/PhysRevD.85.055018}{\emph{Phys. Rev.} {\bf
  D85} (2012) 055018}, [\href{https://arxiv.org/abs/1108.4416}{{\tt
  1108.4416}}].

\bibitem{Maiezza:2015lza}
A.~Maiezza, M.~Nemevšek and F.~Nesti, \emph{{Lepton Number Violation in Higgs
  Decay at LHC}},
  \href{http://dx.doi.org/10.1103/PhysRevLett.115.081802}{\emph{Phys. Rev.
  Lett.} {\bf 115} (2015) 081802},
  [\href{https://arxiv.org/abs/1503.06834}{{\tt 1503.06834}}].

\bibitem{delAguila:2008cj}
F.~del Aguila and J.~A. Aguilar-Saavedra, \emph{{Distinguishing seesaw models
  at LHC with multi-lepton signals}},
  \href{http://dx.doi.org/10.1016/j.nuclphysb.2008.12.029}{\emph{Nucl. Phys.}
  {\bf B813} (2009) 22--90}, [\href{https://arxiv.org/abs/0808.2468}{{\tt
  0808.2468}}].

\bibitem{Atre:2009rg}
A.~Atre, T.~Han, S.~Pascoli and B.~Zhang, \emph{{The Search for Heavy Majorana
  Neutrinos}},
  \href{http://dx.doi.org/10.1088/1126-6708/2009/05/030}{\emph{JHEP} {\bf 05}
  (2009) 030}, [\href{https://arxiv.org/abs/0901.3589}{{\tt 0901.3589}}].

\bibitem{Dev:2016dja}
P.~S.~B. Dev, R.~N. Mohapatra and Y.~Zhang, \emph{{Probing the Higgs Sector of
  the Minimal Left-Right Symmetric Model at Future Hadron Colliders}},
  \href{http://dx.doi.org/10.1007/JHEP05(2016)174}{\emph{JHEP} {\bf 05} (2016)
  174}, [\href{https://arxiv.org/abs/1602.05947}{{\tt 1602.05947}}].

\bibitem{Banerjee:2015gca}
S.~Banerjee, P.~S.~B. Dev, A.~Ibarra, T.~Mandal and M.~Mitra, \emph{{Prospects
  of Heavy Neutrino Searches at Future Lepton Colliders}},
  \href{http://dx.doi.org/10.1103/PhysRevD.92.075002}{\emph{Phys. Rev.} {\bf
  D92} (2015) 075002}, [\href{https://arxiv.org/abs/1503.05491}{{\tt
  1503.05491}}].

\bibitem{Dev:2013wba}
P.~S.~B. Dev, A.~Pilaftsis and U.-k. Yang, \emph{{New Production Mechanism for
  Heavy Neutrinos at the LHC}},
  \href{http://dx.doi.org/10.1103/PhysRevLett.112.081801}{\emph{Phys. Rev.
  Lett.} {\bf 112} (2014) 081801}, [\href{https://arxiv.org/abs/1308.2209}{{\tt
  1308.2209}}].

\bibitem{Das:2012ze}
A.~Das and N.~Okada, \emph{{Inverse seesaw neutrino signatures at the LHC and
  ILC}}, \href{http://dx.doi.org/10.1103/PhysRevD.88.113001}{\emph{Phys. Rev.}
  {\bf D88} (2013) 113001}, [\href{https://arxiv.org/abs/1207.3734}{{\tt
  1207.3734}}].

\bibitem{Das:2016hof}
A.~Das, P.~Konar and S.~Majhi, \emph{{Production of Heavy neutrino in
  next-to-leading order QCD at the LHC and beyond}},
  \href{http://dx.doi.org/10.1007/JHEP06(2016)019}{\emph{JHEP} {\bf 06} (2016)
  019}, [\href{https://arxiv.org/abs/1604.00608}{{\tt 1604.00608}}].

\bibitem{Arhrib:2011uy}
A.~Arhrib, R.~Benbrik, M.~Chabab, G.~Moultaka, M.~C. Peyranere, L.~Rahili
  et~al., \emph{{The Higgs Potential in the Type II Seesaw Model}},
  \href{http://dx.doi.org/10.1103/PhysRevD.84.095005}{\emph{Phys. Rev.} {\bf
  D84} (2011) 095005}, [\href{https://arxiv.org/abs/1105.1925}{{\tt
  1105.1925}}].

\bibitem{Abada:2007ux}
A.~Abada, C.~Biggio, F.~Bonnet, M.~B. Gavela and T.~Hambye, \emph{{Low energy
  effects of neutrino masses}},
  \href{http://dx.doi.org/10.1088/1126-6708/2007/12/061}{\emph{JHEP} {\bf 12}
  (2007) 061}, [\href{https://arxiv.org/abs/0707.4058}{{\tt 0707.4058}}].

\bibitem{Dinh:2013vya}
D.~N. Dinh and S.~T. Petcov, \emph{{Lepton Flavor Violating $\tau$ Decays in
  TeV Scale Type I See-Saw and Higgs Triplet Models}},
  \href{http://dx.doi.org/10.1007/JHEP09(2013)086}{\emph{JHEP} {\bf 09} (2013)
  086}, [\href{https://arxiv.org/abs/1308.4311}{{\tt 1308.4311}}].

\bibitem{Lindner:2016bgg}
M.~Lindner, M.~Platscher and F.~S. Queiroz, \emph{{A Call for New Physics : The
  Muon Anomalous Magnetic Moment and Lepton Flavor Violation}},
  \href{https://arxiv.org/abs/1610.06587}{{\tt 1610.06587}}.

\bibitem{Akeroyd:2009nu}
A.~G. Akeroyd, M.~Aoki and H.~Sugiyama, \emph{{Lepton Flavour Violating Decays
  tau $\to$ anti-l ll and mu $\to$ e gamma in the Higgs Triplet Model}},
  \href{http://dx.doi.org/10.1103/PhysRevD.79.113010}{\emph{Phys. Rev.} {\bf
  D79} (2009) 113010}, [\href{https://arxiv.org/abs/0904.3640}{{\tt
  0904.3640}}].

\bibitem{Chakrabortty:2015zpm}
J.~Chakrabortty, P.~Ghosh, S.~Mondal and T.~Srivastava, \emph{{Reconciling
  (g-2)$_μ$ and charged lepton flavor violating processes through a doubly
  charged scalar}},
  \href{http://dx.doi.org/10.1103/PhysRevD.93.115004}{\emph{Phys. Rev.} {\bf
  D93} (2016) 115004}, [\href{https://arxiv.org/abs/1512.03581}{{\tt
  1512.03581}}].

\bibitem{Esteban:2016qun}
I.~Esteban, M.~C. Gonzalez-Garcia, M.~Maltoni, I.~Martinez-Soler and
  T.~Schwetz, \emph{{Updated fit to three neutrino mixing: exploring the
  accelerator-reactor complementarity}},
  \href{https://arxiv.org/abs/1611.01514}{{\tt 1611.01514}}.

\bibitem{Gonzalez-Garcia:2015qrr}
M.~C. Gonzalez-Garcia, M.~Maltoni and T.~Schwetz, \emph{{Global Analyses of
  Neutrino Oscillation Experiments}},
  \href{http://dx.doi.org/10.1016/j.nuclphysb.2016.02.033}{\emph{Nucl. Phys.}
  {\bf B908} (2016) 199--217}, [\href{https://arxiv.org/abs/1512.06856}{{\tt
  1512.06856}}].

\bibitem{Grimus:2000vj}
W.~Grimus and L.~Lavoura, \emph{{The Seesaw mechanism at arbitrary order:
  Disentangling the small scale from the large scale}},
  \href{http://dx.doi.org/10.1088/1126-6708/2000/11/042}{\emph{JHEP} {\bf 11}
  (2000) 042}, [\href{https://arxiv.org/abs/hep-ph/0008179}{{\tt
  hep-ph/0008179}}].

\bibitem{Deshpande:1990ip}
N.~G. Deshpande, J.~F. Gunion, B.~Kayser and F.~I. Olness, \emph{{Left-right
  symmetric electroweak models with triplet Higgs}},
  \href{http://dx.doi.org/10.1103/PhysRevD.44.837}{\emph{Phys. Rev.} {\bf D44}
  (1991) 837--858}.

\bibitem{Roitgrund:2014zka}
A.~Roitgrund, G.~Eilam and S.~Bar-Shalom, \emph{{Implementation of the
  left-right symmetric model in FeynRules}},
  \href{http://dx.doi.org/10.1016/j.cpc.2015.12.009}{\emph{Comput. Phys.
  Commun.} {\bf 203} (2016) 18--44},
  [\href{https://arxiv.org/abs/1401.3345}{{\tt 1401.3345}}].

\bibitem{Maiezza:2016ybz}
A.~Maiezza, G.~Senjanović and J.~C. Vasquez, \emph{{Higgs Sector of the
  Left-Right Symmetric Theory}},  \href{https://arxiv.org/abs/1612.09146}{{\tt
  1612.09146}}.

\bibitem{Zhang:2007da}
Y.~Zhang, H.~An, X.~Ji and R.~N. Mohapatra, \emph{{General CP Violation in
  Minimal Left-Right Symmetric Model and Constraints on the Right-Handed
  Scale}}, \href{http://dx.doi.org/10.1016/j.nuclphysb.2008.05.019}{\emph{Nucl.
  Phys.} {\bf B802} (2008) 247--279},
  [\href{https://arxiv.org/abs/0712.4218}{{\tt 0712.4218}}].

\bibitem{Maiezza:2014ala}
A.~Maiezza and M.~Nemevšek, \emph{{Strong P invariance, neutron electric
  dipole moment, and minimal left-right parity at LHC}},
  \href{http://dx.doi.org/10.1103/PhysRevD.90.095002}{\emph{Phys. Rev.} {\bf
  D90} (2014) 095002}, [\href{https://arxiv.org/abs/1407.3678}{{\tt
  1407.3678}}].

\bibitem{Bertolini:2014sua}
S.~Bertolini, A.~Maiezza and F.~Nesti, \emph{{Present and Future K and B Meson
  Mixing Constraints on TeV Scale Left-Right Symmetry}},
  \href{http://dx.doi.org/10.1103/PhysRevD.89.095028}{\emph{Phys. Rev.} {\bf
  D89} (2014) 095028}, [\href{https://arxiv.org/abs/1403.7112}{{\tt
  1403.7112}}].

\bibitem{Maiezza:2016bzp}
A.~Maiezza, M.~Nemevšek and F.~Nesti, \emph{{Perturbativity and mass scales in
  the minimal left-right symmetric model}},
  \href{http://dx.doi.org/10.1103/PhysRevD.94.035008}{\emph{Phys. Rev.} {\bf
  D94} (2016) 035008}, [\href{https://arxiv.org/abs/1603.00360}{{\tt
  1603.00360}}].

\bibitem{Chakrabortty:2016wkl}
J.~Chakrabortty, J.~Gluza, T.~Jelinski and T.~Srivastava, \emph{{Theoretical
  constraints on masses of heavy particles in Left-Right Symmetric Models}},
  \href{http://dx.doi.org/10.1016/j.physletb.2016.05.092}{\emph{Phys. Lett.}
  {\bf B759} (2016) 361--368}, [\href{https://arxiv.org/abs/1604.06987}{{\tt
  1604.06987}}].

\bibitem{Aad:2015iea}
{\scshape ATLAS} collaboration, G.~Aad et~al., \emph{{Summary of the searches
  for squarks and gluinos using $ \sqrt{s}=8 $ TeV pp collisions with the ATLAS
  experiment at the LHC}},
  \href{http://dx.doi.org/10.1007/JHEP10(2015)054}{\emph{JHEP} {\bf 10} (2015)
  054}, [\href{https://arxiv.org/abs/1507.05525}{{\tt 1507.05525}}].

\bibitem{Aad:2015pfx}
{\scshape ATLAS} collaboration, G.~Aad et~al., \emph{{ATLAS Run 1 searches for
  direct pair production of third-generation squarks at the Large Hadron
  Collider}}, \href{http://dx.doi.org/10.1140/epjc/s10052-015-3726-9,
  10.1140/epjc/s10052-016-3935-x}{\emph{Eur. Phys. J.} {\bf C75} (2015) 510},
  [\href{https://arxiv.org/abs/1506.08616}{{\tt 1506.08616}}].

\bibitem{Khachatryan:2016xdt}
{\scshape CMS} collaboration, V.~Khachatryan et~al., \emph{{Search for
  supersymmetry in events with one lepton and multiple jets in proton-proton
  collisions at sqrt(s) = 13 TeV}}, {\emph{Submitted to: Phys. Rev. D} (2016)
  }, [\href{https://arxiv.org/abs/1609.09386}{{\tt 1609.09386}}].

\bibitem{Khachatryan:2016epu}
{\scshape CMS} collaboration, V.~Khachatryan et~al., \emph{{Inclusive search
  for supersymmetry using razor variables in pp collisions at sqrt(s) = 13
  TeV}}, {\emph{Submitted to: Phys. Rev. D} (2016) },
  [\href{https://arxiv.org/abs/1609.07658}{{\tt 1609.07658}}].

\bibitem{Dittmaier:2007uw}
S.~Dittmaier, G.~Hiller, T.~Plehn and M.~Spannowsky, \emph{{Charged-Higgs
  Collider Signals with or without Flavor}},
  \href{http://dx.doi.org/10.1103/PhysRevD.77.115001}{\emph{Phys. Rev.} {\bf
  D77} (2008) 115001}, [\href{https://arxiv.org/abs/0708.0940}{{\tt
  0708.0940}}].

\end{thebibliography}\endgroup
\end{document}